\documentclass[]{aa} 

\usepackage{graphicx}
\usepackage{txfonts}
\usepackage{natbib}
\usepackage{longtable}
\usepackage{lscape}
\usepackage{color}

\providecommand{\sorthelp}[1]{}

\newcommand{\Planck}{{\it Planck }}
\newcommand{\Herschel}{{\it Herschel }}
\newcommand{\Spitzer}{{\it Spitzer }}

\begin{document}
 
\title{
{Dust emission and extinction in the Orion OMC-3 cloud}
}

\author{
  {Mika Juvela\inst{1}}
  \and
      {Nathalie Ysard\inst{2,3}}}

\institute{
Department of Physics, P.O.Box 64, FI-00014, University of Helsinki,
Finland, {\em mika.juvela@helsinki.fi}
\and
Universit\'e de Toulouse, UPS-OMP, IRAP, F-31028 Toulouse cedex 4, France
\and
Universit\'e Paris-Saclay, CNRS, Institut d'Astrophysique Spatiale, 91405, Orsay,
France
}

\authorrunning{M. Juvela et al.}

\date{Received September 15, 1996; accepted March 16, 1997}

\abstract { 
  Dust is an important tracer of the structure of interstellar clouds, as well as a
  central factor in the thermal balance and chemistry of the clouds. Our knowledge of the
  dust properties is nevertheless incomplete, especially regarding the dense star-forming
  clouds.
} 
{
  The aim is to study dust evolution in the Orion Molecular Cloud 3 (OMC-3) and how 
  uncertainty regarding dust properties affects estimates of the radiation field and the cloud
  mass.
}
{
  We constructed three-dimensional radiative transfer (RT) models to fit the far-infrared (FIR) observations of dust emission in the OMC-3 field and used near-infrared (NIR) extinction
  measurements as additional constraints. We examined fits to the dense star-forming filaments and
  to the surrounding cloud, including some tests with spatial dust property variations.
}
{
  The 160-250\,$\mu$m observations of dust emission could be fitted moderately well with any of the
  dust models tested, but few models are consistent with the measured NIR extinction. The best
  match to observations is found with dust models such as the THEMIS model of large porous grains,
  with or without ice mantles, and with mean grain sizes up to $\sim 0.3\,\mu$m. The flattening of
  the NIR extinction curve excludes larger grain sizes, except possibly in the central
  ridge. Compared to models of lower column density clouds, the results were relatively insensitive
  to the line-of-sight (LOS) cloud size and the spectral shape of the heating radiation field.  In addition, the
  effect of embedded stars remained very localised in OMC-3.
}
{ 
The results suggest that the dust in the OMC-3 region is evolved with a grain of
average size $a=0.1-0.3\,\mu$m, potentially with ice mantles.
}


\maketitle

\section{Introduction} \label{sect:intro}

Space-borne missions have gathered comprehensive observations from mid-infrared (MIR) to
far-infrared (FIR) wavelengths and even to the radio regime. These include both all-sky
observations, such as in the {\it WISE} \citep[3-22\,$\mu$m;][]{Wright2010}, {\it AKARI}
\citep[2-160\,$\mu$m;][]{Murakami2007}, and \Planck \citep[$\lambda\ge 350\,\mu{\rm
    m}$;][]{Tauber2010} surveys, as well as more targeted observations with the \Spitzer
\citep[$\lambda=3.6-160\,\mu$m;][]{Werner2004}, and \Herschel \citep[$\lambda=70-500\,\mu{\rm
    m}$;][]{Pilbratt2010} satellites. These provide a good starting point for investigations of 
dust evolution in star-forming clouds. The MIR and FIR parts of the spectrum trace the thermal dust
emission but from different components of the overall dust populations. The FIR spectrum is
produced by large grains that are mostly in radiative equilibrium with the local radiation
field. In contrast, the MIR emission comes from stochastically heated smaller grains and is
therefore more sensitive to the lower end of the dust size distribution and the thermal properties
of those grains.

Studies of FIR thermal dust emission have revealed variations in the large-grain properties
over the whole sky \citep{Planck2014_allsky_model}, especially within the transitions from atomic
to molecular clouds and further to the dense clumps that are more immediately connected to the
present star formation in the Galaxy \citep{Lagache1998, Flagey2009, Cambresy2001, Stepnik2003,
  Roy2013}. The dust opacity spectral index $\beta$ and the absolute value of the opacity itself,
$\kappa$, vary spatially and as a function of wavelength. The first evidence of this evolution came
from balloon-borne FIR and millimetre wavelength observations \citep{Bernard1999, Dupac2003,
  Desert2008, Paradis2009, Martin2012}, and the variations have been studied in more detail with
satellite data, especially with \Herschel and \Planck \citep{Paradis2010, Planck2014_allsky_dust,
  Planck2015_freqdep}.  The results indicate that the dust FIR spectrum may become steeper (with
higher $\beta$) and the dust opacity may increase towards dense clumps and cores
\citep{Stepnik2003,GCC-V,GCC-VI,Scibelli2023}, while the spectrum generally flattens towards longer
       millimetre wavelengths
      \citep{Reach1995,PER_XVII,PER_XIX,Planck2016_XXIX,Paradis2012_500um}. These results indicate changes
      in both the sizes and optical properties of the grains. Ground-based instruments
      contribute to the studies with observations of higher angular resolution, probing the
      internal structure of starless and protostellar cores in the nearby clouds. On the other
      hand, the elimination of atmospheric effects leads to limited sensitivity to extended
      emission \cite[e.g.][]{Bianchi2003,Sadavoy2013,Bracco2017,Juvela_2018_pilot}, which
      complicates the combination of space-borne and ground-based observations
      \citep[e.g.][]{Schuller2021,Mannfors2025}.

The observational constraints on the dust evolution are stronger if the FIR observations of dust
emission are complemented with data from shorter wavelengths. When a high column density cloud is
seen against a bright background, it may be possible to use mid-infrared (MIR) absorptions to map
the cloud structure. This is the case for massive infrared dark clouds (IRDCs) that are observed
against the brighter Galactic disk \citep{Simon2006a,ButlerTan2012,KainulainenTan2013}. The method
requires measurements (or estimates) of the absolute surface brightness of the background sky and,
depending on the wavelength, may have additional uncertainty due to variations in the local heating
(e.g. embedded protostars) and potential contribution from light scattering
\citep{Juvela2023_OMC3}.

The NIR extinction is measured more commonly using background point sources. If embedded sources
are few or can be excluded, the method is insensitive to local heating and thus applicable even in
active star-forming regions. The NIR reddening of stars (e.g. measured with ground-based
observations in the J, H, and Ks bands) can be converted to a continuous extinction map with
methods such as NICER \citep{Lombardi2001} and NICEST \citep{Lombardi2009}. These combine the
measured colour excesses and smooth the evidence provided by the individual stars, typically down
to a resolution of some arcminutes. The resolution is limited mainly by the number density of the
background sources. Apart from the intrinsic colours of the stars (typically estimated using a
nearby extinction-free reference field), the extinction estimates depend only on the assumed shape
of the NIR extinction curve. The extinction curve is not constant, but has shown relatively
little variation between clouds of different density \citep{Indebetouw2005, Wang2014}. In contrast,
the FIR to NIR dust opacity ratio has been observed to increase by a factor of a few from diffuse
clouds to dense clumps. This is clear evidence of dust evolution and is thus mainly attributed to
the increase in FIR dust opacity
\citep{Martin2012,Roy2013,Ysard2013,GCC-V,Juvela_L1642,Juvela2024_Taurus}.

The interpretation of the FIR data in dense clouds is complicated by the strong temperature
gradients that exist between the warm surface layers and the colder interiors of dense filaments
and starless cores. One way to take these temperature variations into account, at least
approximately, is to use three-dimensional radiative transfer (RT) modelling, where the dust
temperatures follow self-consistently from the assumptions of the density field, the dust
properties, and the radiation sources. Modelling can still have significant uncertainties due to
the degeneracies between these factors, especially because the LOS cloud structure
and the local radiation field spectra are not directly constrained by observations
\citep{Juvela2024_Taurus}. Models have often concentrated on the combination of FIR emission and
NIR extinction \citep[e.g.][]{Stepnik2003,Ysard2013,GCC-V}, because sensitive measurements of MIR
surface brightness are more rare. The light scattering would also provide more information on the
grain properties, but is mostly useful when the model geometry is simple and the intensity of the
scattered signal has not yet saturated (i.e. for optical depths $\tau(\lambda)<1$).  The NIR
scattering (the so-called cloudshine) has been used in only a few cases \citep{Juvela2012_CrA,
  Saajasto2021,Juvela_L1642}, but the MIR scattering (the so-called coreshine) has been
investigated for a larger sample of dense cores imaged with \Spitzer \citep[e.g.][]{Steinacker2010,
  Pagani2010, Andersen2014, Steinacker2015, Lefevre2016}.

In this paper, we use NIR extinction and FIR dust emission to study dust properties in the Orion
Molecular Cloud 3 (OMC-3). Orion is an interesting target as the nearest example of a high-mass
star-forming region, with a distance of $\sim$400\,pc \citep{Grossschedl2018}. In OMC-3 the main
structure is a massive filament, part of the northern end of the Orion integral-shaped filament
\citep{Kainulainen2017,Hacar2018,Schuller2021}.  OMC-3 thus provides an important point of
comparison for previously studied filamentary low-mass star-forming clouds and the more massive and
more distant IRDCs. Compared to other parts of the OMC, the star-formation activity is still
relatively low in OMC-3. This simplifies the modelling, since, at least for the central part of the
field, dust is mainly heated by the external radiation field.
We concentrate on the modelling of the FIR large-grain emission, using the extinction measurements
afterwards to test the accuracy of the model predictions at NIR wavelengths. We test
alternative dust models to see which of them are compatible with the OMC-3 observations. The
complete modelling of MIR-FIR observations is deferred to a future paper in which we will address the
partly separate questions of small-grain emission, MIR absorption, and NIR-MIR scattering.

The contents of the paper are the following. We present the observational data in
Sect.~\ref{sect:observations}. The three-dimensional RT modelling methods and the dust models
employed are discussed in Sect.\ref{sect:methods}, and the results of the optimised RT models are
presented in Sect.~\ref{sect:results}. We discuss the results in Sect.~\ref{sect:discussion},
before listing our final conclusions in Sect.~\ref{sect:conclusions}.

\section{Observational data} \label{sect:observations}

\subsection{Observations of extended emission} \label{sect:obsI}

The Orion Molecular Cloud 3 was mapped with the \Herschel telescope as part of the Gould Belt Survey \citep[PI
  Ph. Andre;][]{Andre2010}. The 250\,$\mu$m, 350\,$\mu$m, and 500\,$\mu$m observations with the
SPIRE instrument were done in parallel mode (observation IDs 1342218967 and 1342218968), together
with 70\,$\mu$m and 160\,$\mu$m observations with the PACS instrument. The relative calibration
accuracy of the SPIRE observations is expected to be better than
2\%\footnote{https://www.cosmos.esa.int/web/herschel/spire-overview}, and for the PACS data we
assume 4\% uncertainty relative to SPIRE.

The resolution of the SPIRE observations is about 18, 26, and 37 arcsec, at
250\,$\mu$m, 350\,$\mu$m, and 500\,$\mu$m, respectively. For the modelling, the
maps were degraded to lower resolution using the convolution kernels of
\cite{Aniano2011}, with the final resolution corresponding to Gaussian beams
with FWHM values of 30$\arcsec$, 30$\arcsec$, and 41$\arcsec$, respectively. The
size of the map pixels was set to 4$\arcsec$, which also corresponds
to the spatial discretisation of the three-dimensional models
(Sect.~\ref{sect:density}). The nominal resolution of the fitted models was
30$\arcsec$ (as described in Sect.~\ref{sect:methods}), and the PACS 70\,$\mu$m
and 160\,$\mu$m maps were thus also convolved to this resolution. The lower
spatial resolution decreased the run times of the model optimisation, with only
minor loss of information. The 500\,$\mu$m band remains important for
constraining the spectral index (and therefore also the temperatures) and was
 used near the original resolution.

\subsubsection{Colour corrections}

 To simplify the comparison with models, observed surface brightness maps were colour corrected
  to monochromatic values at the nominal wavelengths. For the SPIRE data, the correction was based
  on modified black-body (MBB) spectra, $I_{\nu} \propto B_{\nu}(T)\times \nu^{\beta}$,
  where the temperatures were obtained from MBB fits to the SPIRE data at 41$\arcsec$ angular
  resolution. The fits used a constant dust opacity spectral index of $\beta=1.8$
  (cf. Sect.~\ref{sect:MBB}) and took into account colour corrections  in the fits.  The
SPIRE colour corrections are only a couple of per cent and also change slowly as functions of
temperature and spectral index. The PACS 160\,$\mu$m data were colour corrected for the same MBB
spectra. We do not use the 70\,$\mu$m data in this paper, because these can already have a
significant contribution from stochastically heated grains.

\subsubsection{Background subtraction} \label{sect:bgsub}

The initial modelling relied on the SPIRE surface brightness data. In the Herschel Science
Archive\footnote{https://archives.esac.esa.int/hsa}, the zero points of the SPIRE intensity scales
have already been adjusted by comparing the maps with \Planck absolute surface brightness
measurements. However, the RT models represent only part of the OMC-3 cloud, also in the LOS
direction, and one has to subtract from the maps the extended emission that is thought to originate
outside this volume. Background subtraction was done by subtracting the mean value under a
$FWHM=1\arcmin$ Gaussian beam that was placed at the first reference position of  Ra=5:34:46.1
and Dec=-5:01:59.3 (J2000). In the following, $I_{\nu}$ refers to these background-subtracted
intensities.
 
\begin{figure}
\sidecaption
\includegraphics[width=8.8cm]{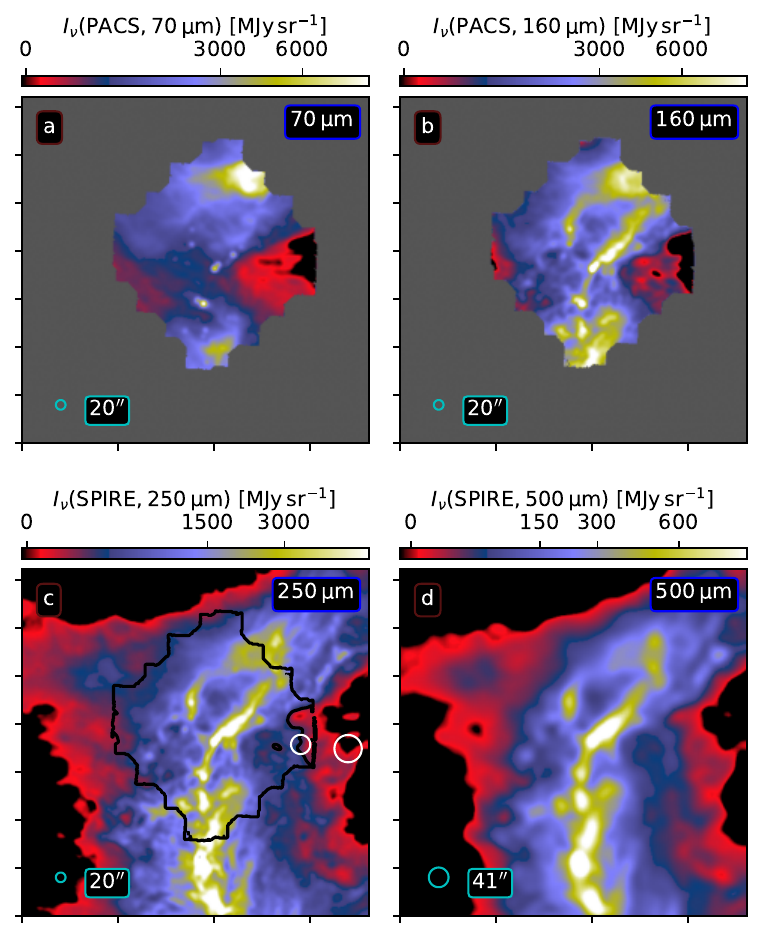}
\caption{
Examples of OMC-3 observations. The maps show background-subtracted surface brightness values
(non-linear colour scale) smoothed to the resolution used in the model fits. The plotted area is
identical to previous figures. Frame c shows the outline of the PACS coverage and the reference
areas for the SPIRE background subtraction (larger white circle) and for the adjustment of the PACS
zero level (smaller white circle).
}
\label{fig:plot_map}
\end{figure}

The first reference position is outside the area covered by PACS observations
(Fig.~\ref{fig:plot_map}). Therefore, we used a second reference position at Ra=5:34:59.3 and
Dec=-5:01:41.9 (J2000) that is located inside the PACS maps. The PACS zero point was adjusted so
that in this position the corrected PACS map matches the extrapolation from the already
background-subtracted SPIRE data. The extrapolation was done using an MBB spectrum that was fitted
to the SPIRE bands, and for the comparison the surface brightness values were averaged over a
1$\arcmin$ beam. At the second reference position, the 160\,$\mu$m surface brightness is less than
$\sim$10\% of the values in the main filament. The error in the procedure used to set the PACS zero
level is a fraction of this number (e.g. 1\,K error in temperature plus 0.2 unit error in $\beta$
would result in a less than 20\% error in the extrapolation and thus would have less than 2\% effect on the
final values in the filament). Therefore, we assume that the 4\% relative calibration uncertainty
of PACS data also covers the PACS zero point uncertainty in the main filament (relative to
SPIRE). Figure~\ref{fig:plot_map} shows the final background-subtracted maps at 160, 250, and
500\,$\mu$m. The figure also shows for reference the 70\,$\mu$m map , which, however, is not used in
the subsequent analysis.

\subsection{Cloud masks} \label{sect:masks}

\begin{figure}
\includegraphics[width=8.8cm]{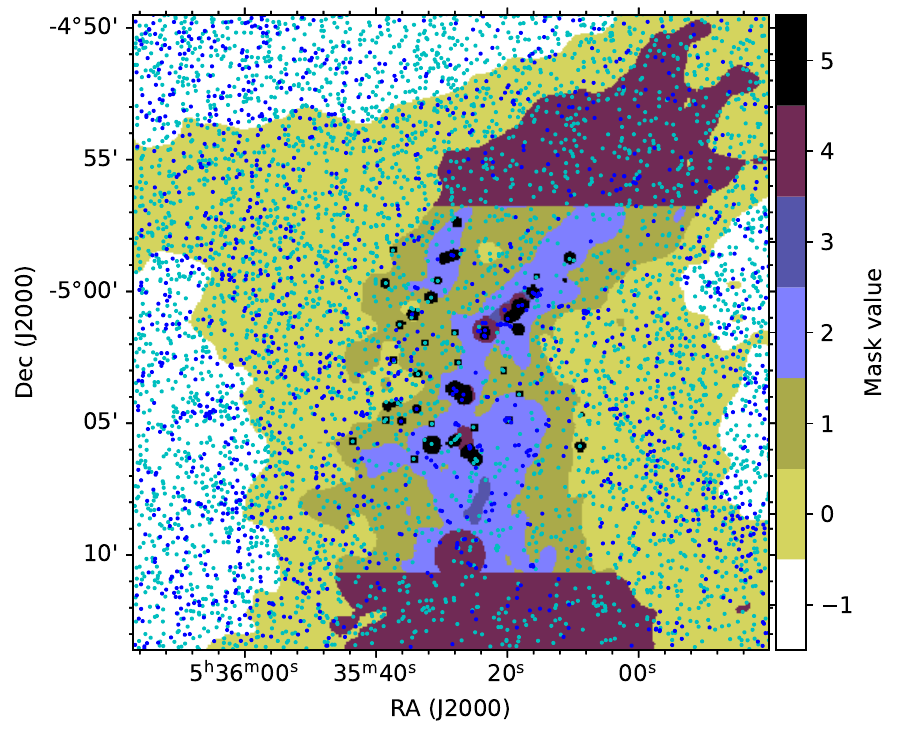}
\caption{
Masks for the OMC-3 field. The levels correspond to areas with low emission ($Q=-1$), extended
cloud ($Q=0$), filament ($Q=1$), ridge ($Q=2$), bright parts of the ridge ($Q=3$), FIR point
sources and high-surface brightness areas affected by local heating ($Q=4$), and MIR point sources
($Q=5$). The small dots correspond to the NIR star-like (cyan) and potentially extended (blue)
sources that are the basis of the NIR extinction map \citep{Meingast2016}.
}
\label{fig:make_masks}
\end{figure}

We used infrared maps to create a mask that was used in the model optimisation and in the
comparison of the model predictions and observations. The mask has six levels, as shown in
Fig.~\ref{fig:make_masks}. The value $Q=-1$ corresponds to regions where the background-subtracted
FIR surface brightness is close to zero or negative ($I_{\nu} \rm (350\,\mu m)<1 \, MJy\,sr^{-1}$)
and are excluded from the analysis. The values $Q=0-2$ correspond, respectively, to regions that we
refer to as the diffuse background ($1 \, {\rm MJy\,sr^{-1}} < I_{\nu} {\rm (350\,\mu m)<190 \,
  MJy\,sr^{-1}}$), the filament region ($190 \, {\rm MJy\,sr^{-1}} < I_{\nu} {\rm (350\,\mu m)<600
  \, MJy\,sr^{-1}}$), and the ridge regions ($600 \, {\rm MJy\,sr^{-1}} < I_{\nu} {\rm (350\,\mu
  m)<2500 \, MJy\,sr^{-1}}$). We further separated the brightest regions of the ridge as $Q=3$
($I_{\nu} \rm (350\,\mu m)>2500 \, MJy\,sr^{-1}$) where the modelling could be unreliable due to
potential unresolved internal radiation sources.

The value $Q=4$ is reserved for resolved FIR point sources (masked by hand), but also excludes the
southern and northern ends of the filament, where the surface brightness increases due to embedded
sources and luminous stars outside the map. These cuts are made along lines of constant
declination, which correspond roughly to the contour $T_{\rm dust}=25$\,K from the MBB temperature
map. However, the mask is more conservative, excluding all regions far from the field centre.  It
excludes especially the southern parts, where high column densities make the temperatures low in
spite of the increasing strength of the external radiation.  Strong MIR point sources were masked
by hand and given a mask value of $Q=5$, even when there is no direct evidence that these sources
affect FIR emission at the 30\,$\arcsec$ scale. All regions with $Q\ge3$ were excluded from the
analysis.

\subsection{NIR extinction} \label{sect:extinction}

We used the  J, H, and K$_{\rm S}$ band NIR observations from the Vienna survey in Orion
\citep[VISION;][]{Meingast2016} and also their NIR extinction map, made using the NICER
method \citep{Lombardi2001}, and the \citet{Indebetouw2005} extinction curve. The map has an angular
resolution of $FWHM=1\arcmin$ beam.

\begin{figure}
\includegraphics[width=9cm]{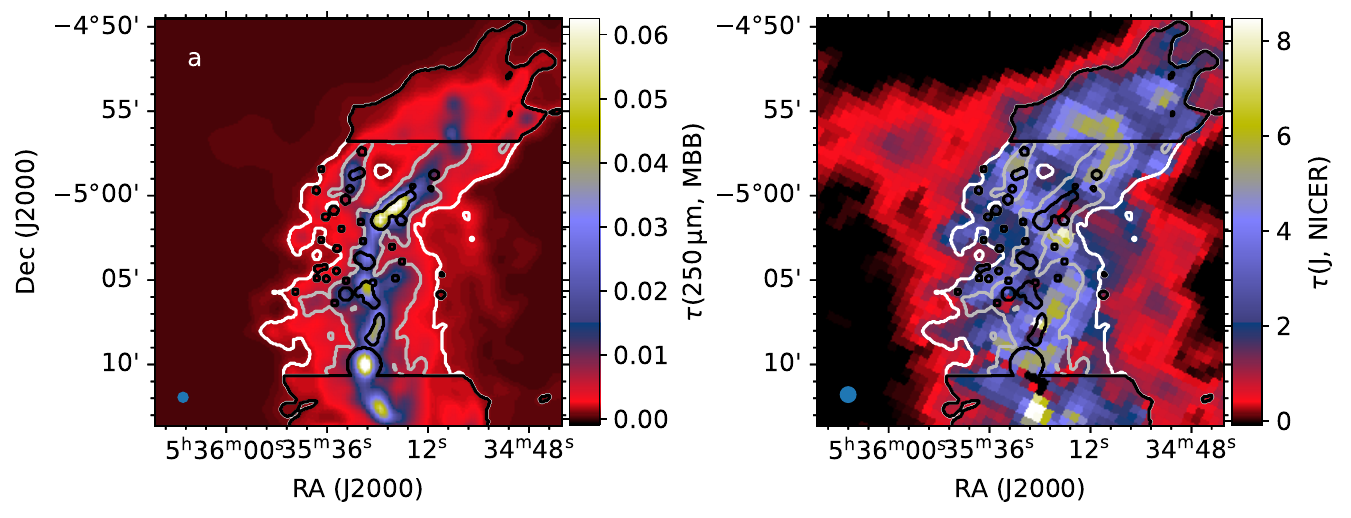}
\caption{
Comparison of  $\tau \rm(250\,\mu m)$ optical depth map derived from \Herschel observations at
41$\arcsec$ resolution (frame a) and NICER extinction map of \citet{Meingast2016} at $1\arcmin$
resolution (frame b). The white and grey contours show, respectively, the outlines of the filament
($Q=1$) and the ridge ($Q=2$) regions. The black contours separate other high column density areas
that are excluded from the analysis ($Q>2$).
}
\label{fig:plot_extinction}
\end{figure}

Figure~\ref{fig:plot_extinction} shows a map of the 250\,$\mu$m optical depth (estimated from
\Herschel data at 41$\arcsec$ resolution, cf.  Sect.~\ref{sect:MBB}) and the NICER extinction map
from \citet{Meingast2016}.  The original NICER map was resampled to equatorial coordinates using
smaller pixels and no interpolation, retaining the original nominal $\sim$1$\arcmin$
resolution. Although the filament region is clearly visible on both maps, the NICER map shows less
correlation with the regions of highest {\it Herschel} column densities (e.g. along the $Q=2$
ridge).

The area of Fig.~\ref{fig:make_masks} contains 3141 sources detected in all three NIR bands. These
can even be used as individual probes of the extinction, with much greater statistical uncertainty
but without the bias introduced by the spatial averaging of the data that show gradients in the
stellar number density. The density of background stars decreases from the diffuse background (1126
sources in the $Q=0$ area) to the filament ($Q=1$ with 210 sources) and further to the ridge ($Q=2$
with 70 sources). In the ridge the sampling is also non-uniform and the probability for embedded
and foreground sources increases. After subtracting the average extinction in the $Q=0$ area (for
the individual stars and the NICER map separately) and rejecting the individual stars with negative
extinction (extinction below the average in the $Q=0$ area, such as possible foreground sources),
the median ratio between the extinction of the individual stars and the NICER values, read at the
same positions, is 0.95 in the $Q=1$ and 1.14 in the $Q=2$ area. The NICER map could be biased
towards the ridge, when stars are detected preferentially towards the lower column density part of
each resolution element. However, comparison with the median value of individual stars in the $Q=2$
area shows a difference of less than $\sim$10\%. The bias could be larger close to the unresolved
density peaks, but these fall mainly inside the already excluded $Q>2$ mask
(Fig.~\ref{fig:make_masks}).  We used the NICER $A({\rm J})$ extinction map for visual inspection
(Fig.~\ref{fig:plot_extinction}b, 1$\arcmin$ resolution). When the extinction of an RT model was
compared to observations (and especially when estimates need to be recalculated for the extinction
curve of the dust model in question), we used the median $A({\rm J})$ of the individual stars
inside a given mask region.  For consistency with modelling that used background-subtracted FIR
maps and to take into account the lower resolution and the large relative noise of the extinction
estimates, background subtraction was done by subtracting the median extinction in the $Q$=-1
area. That includes the SPIRE reference position and, based on Fig.~\ref{fig:plot_extinction},
already has an average extinction close to zero.

\section{Methods}  \label{sect:methods}

The RT modelling started with the definition of the initial three-dimensional density field,
the radiation sources, and the dust model. The column densities and the strength of the radiation
sources were then modified iteratively to provide the best match to the FIR data. The fit procedure
was similar to that used in \citet{Juvela2024_Taurus}.

\subsection{Modified black-body fits} \label{sect:MBB}

MBB fits were used to initialise the column densities of the cloud models. Assuming a single dust
temperature along the full LOS and optically thin emission, the dust (colour) temperature $T_{\rm
  d}$ was obtained by fitting multi-frequency observations with
\begin{equation}
I_{\nu} \propto  B_{\nu}(T_{\rm d}) \times \nu^{\beta},
\end{equation}
with Planck law $B_{\nu}$. The optical depth at a given frequency $\nu$ is then
\begin{equation}
\tau_{\nu} = I_{\nu} / B_{\nu}(T_{\rm d}).
\end{equation}
The optical depth is $\tau_{\nu}=\kappa_{\nu} \Sigma$, where $\kappa_{\nu}$ is the mass absorption
coefficient. $\Sigma$ is the mass surface density, which can be further converted, for example, to
the hydrogen column density. The opacity was assumed to follow a power law,
$\kappa_{\nu} \propto \nu^{\beta}$, where the opacity spectral index was set by default to $\beta$=1.8. The surface
brightness data were convolved to a common resolution for the MBB calculations, which also made it
possible to do these calculations independently for each map pixel.

\subsection{Initial density field}  \label{sect:density}

The model clouds were discretised onto a three-dimensional Cartesian grid of 362$^3$ cells, and the
calculations correspondingly produced surface brightness maps of $362 \times 362$ pixels with
$4\arcsec$ pixels. The maps cover a projected area of $24.1\arcmin \times 24.1\arcmin$, matching
pixel by pixel the resampled observations described in Sect.~\ref{sect:observations}.

The initial density fields were created by combining the column density maps obtained from the MBB
analysis of {\it Herschel} FIR maps with the assumption of the LOS density profile. The initial
plane-of-sky (POS) distribution (i.e. the column densities) did not affect the final result because all column densities were updated during the model optimisation. However, the
cloud shape and extent in the LOS direction was potentially an important parameter that was not well
constrained by observations and also remained unaltered during the optimisation.

The initial density distribution was a combination of Gaussian and Plummer profiles. The Plummer
density profile is \citep{Arzoumanian2011}
\begin{equation}
n(r_{\rm 3D}) = \frac{n_0}{[1+ (r_{\rm 3D}/R_{\rm 0})^2]^{p/2}},
\end{equation}
where $r_{\rm 3D}$ is measured radially (in local cylinder symmetry) from the filament axes traced
by eye (Fig.~\ref{fig:combo}).  Here, $n_0$ is the peak density, $R_{\rm 0}$ the size of the central
flat part, and $p$ the power law index that defines the asymptotic behaviour at large distances.
The Gaussian profiles are a function of the LOS coordinate $r_{\rm LOS}$, which is measured from
the LOS central plane of the model. The peak of the Gaussian component was scaled to $n_0^{\rm G}$,
which was set equal to 1\% or 10\% of the peak Plummer value $n_0$. The final density of a cell is
the maximum of these two components,
\begin{equation}
  n = {\rm max}\{ {\rm Plummer}(r_{\rm 3D}; n_0, R_0, p), {\rm Gauss}(r_{\rm LOS}; n_0^{\rm G},
  FWHM) \}.
\end{equation}
 The Gaussian component was initially constant over the sky, the final column density
  structure then resulting from the model optimisation.

\begin{figure}
\includegraphics[width=9cm]{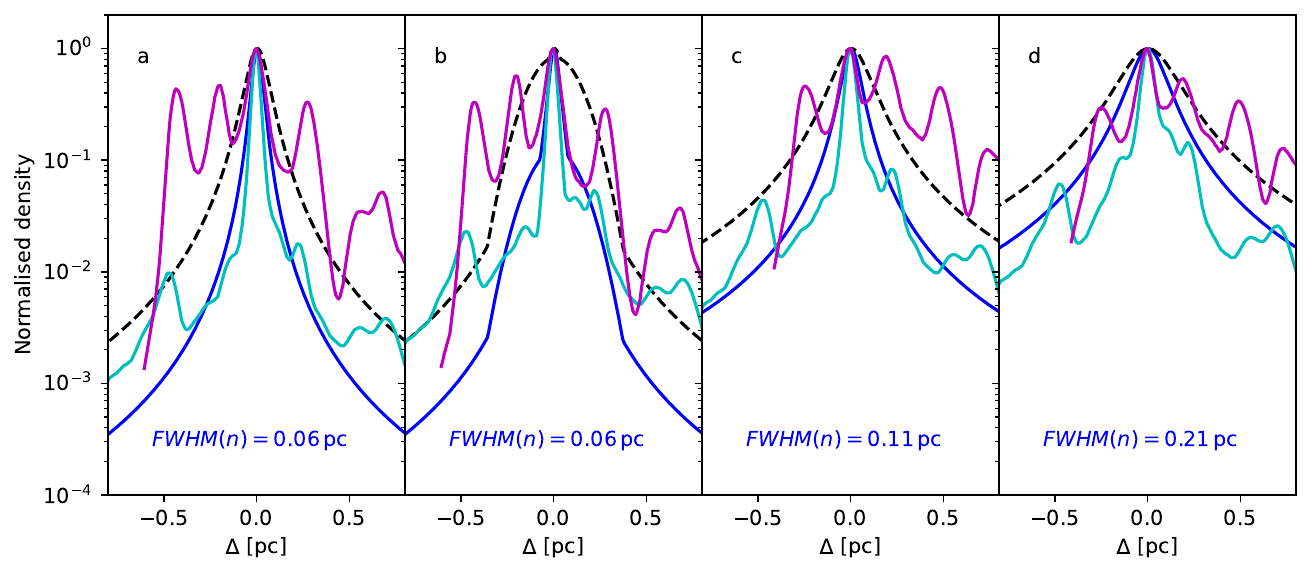}
\caption{ 
  Normalised density profiles (blue lines) for cloud models A-D, for LOS towards filament
  (position indicated in Fig.~\ref{fig:combo}). Examples of profiles at FWHM distance from the
  filament spine are plotted with dashed black lines. The density profiles for the other orthogonal
  directions, along constant right ascension (magenta lines) and constant declination (cyan), are
  for the model A with THEMIS dust.
  The quoted $FWHM$ values correspond to the blue curve.
}
\label{fig:plot_LOS_profiles}
\end{figure}

We used four alternative initial density fields. Their parameters are listed in
Table~\ref{table:clouds}, and Fig.~\ref{fig:plot_LOS_profiles} shows the corresponding relative
densities for a LOS towards a filament. The Gaussian component determines the cloud shape at large
projected distances from the filaments, especially in model B. The Plummer component dominates
close to the filament spines (Fig.~\ref{fig:combo}) and remains important for the whole LOS towards
the filaments. The model optimisation (Sect.~\ref{sect:fits}) changes the model column densities
but leaves the shape of the LOS density profiles unchanged. The final models are also independent
of the initial absolute scaling of the density values.

\begin{table}
\caption{Parameters for cloud density field in  A-C cloud models.}
\begin{center}
\begin{tabular}{llllll}
\hline
&  Gauss         &   Ratio                & Plummer      &       &         \\
Name  &  $FWHM$  &   $n_0^{\rm G}/n_0$    & $R_{\rm 0}$  &  $p$  &  $FWHM$ \\
&        (pc)    &                        & (pc)         &       &  (pc)   \\
\hline
A  &   0.3          &   0.01          &  0.03        &  2.5  &  0.074  \\  
B  &   0.3          &   0.1           &  0.03        &  2.5  &  0.074  \\  
C  &   0.4          &   0.1           &  0.05        &  2.0  &  0.173  \\  
D  &   0.4          &   0.1           &  0.1         &  2.0  &  0.346  \\  
\hline
\end{tabular}
\end{center}
\label{table:clouds}
\end{table}

\subsection{Radiation field}  \label{sect:radiation}

The model clouds are heated by an isotropic external radiation field that was assumed by default to
have the spectral shape of a normal interstellar radiation field (ISRF) \citep[ISRF;][]{Mathis1983}.
The actual radiation field in OMC-3 is affected by nearby stars (especially B stars both south and
north of the central OMC-3) and by the dust attenuation caused by the rest of the cloud around the
modelled sub-volume.  We therefore also considered alternative spectral shapes for the incoming
radiation. These correspond to a fixed value for an external extinction $A^{\rm ext}_{\rm V}$ and
the extinction curve of the CMM dust model (see Sect.~\ref{sect:dust}). Based on {\it Herschel}
observations, the external cloud layer should correspond to $A^{\rm ext}_{\rm V}\sim1$\,mag or
less. However, since the absolute level of the radiation field is a separate free parameter,
$A^{\rm ext}_{\rm V}$ should be considered purely as a parameter for the spectral shape of the
incoming radiation. Thus, in addition to positive values of $A^{\rm ext}_{\rm V}$=0.5-2\,mag, we
also tested $A^{\rm ext}_{\rm V}=-0.5$\,mag, which corresponds to a harder radiation field
spectrum. The absolute scaling of the radiation field was left as a free parameter
(Sect.~\ref{sect:fits}).

\begin{figure}
\includegraphics[width=8.8cm]{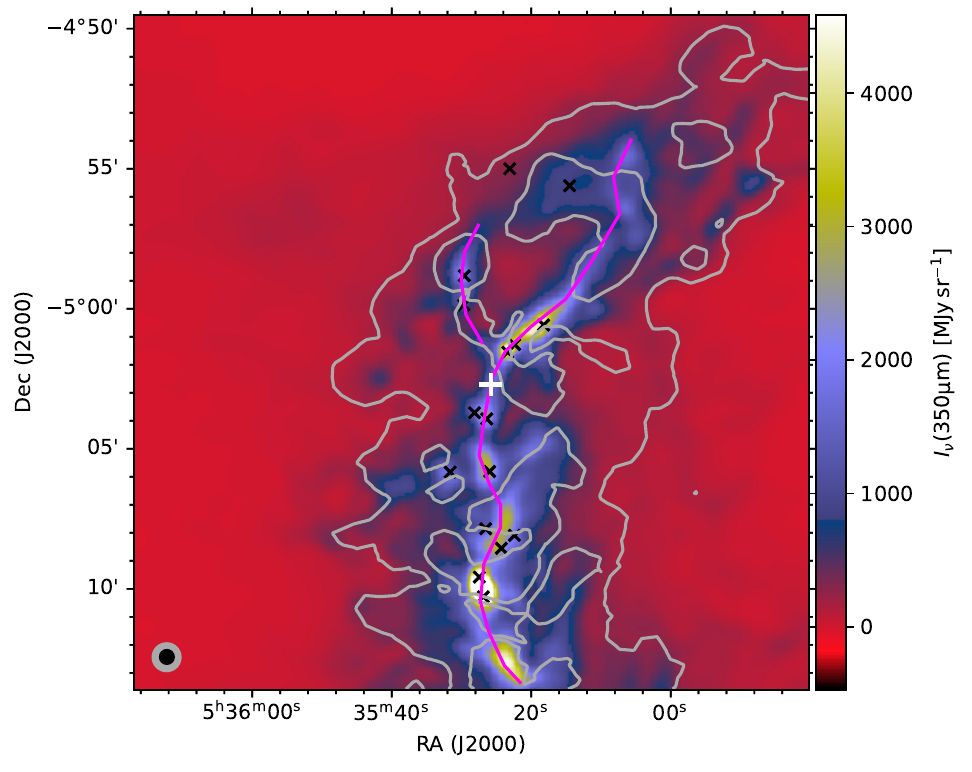}
\caption{
  OMC-3 surface brightness at 350\,$\mu$m (colour scale). The grey contours correspond to NIR
  extinction $A_{\rm J}$=2.5\,mag and 5\,mag. The black crosses show the locations of potential
  embedded radiation sources that were included in some models. The magenta lines trace the
  filament spines that were used to generate the initial model density distributions.  The
  white plus sign shows the position for the LOS density profiles in
  Fig.~\ref{fig:plot_LOS_profiles}.
}
\label{fig:combo}
\end{figure}

Some tests were made by including 16 point sources to represent the potential internal heating
(Fig.~\ref{fig:combo}). The positions were selected as a combination of the sources at 24\,$\mu$m
or 70\,$\mu$m that appear to be related to some increase in the local FIR surface brightness or in the
160\,$\mu$m/250\,$\mu$m ratio (i.e. colour temperature). The sources were modelled as 7000\,K black
bodies, and the source luminosities were left as free parameters that were optimised based on the
surface brightness within an annulus covering the distances $r=10-40\arcsec$ from each
  source. The sources reside in areas with mask values $Q=3-4$ and thus do not directly affect the
$k_{\rm ISRF}$ parameter. The sources were placed at the cloud centre in the LOS direction, in the
region of highest volume density. Due to the low model resolutions, the black-body sources
represent emission escaping a single model cell and not directly the radiation of an unresolved
stellar or protostellar source.

In the northern and southern parts of the field, the emission becomes dominated by sources close to
and beyond the map edges, and those parts of the filament (with $Q=4$) are thus excluded from the
analysis. The RT models are more accurate towards the centre of the field, where the assumption of
an isotropic external radiation field is more appropriate. Appendix~\ref{app:radiation_field}
  shows examples of dust emission spectra at central declinations. These indicate only modest
  variations in the colour temperature of the large-grain emission.

\subsection{Dust models} \label{sect:dust}

Our goal was to find out to what extent the dust properties in OMC-3 can be constrained with the
available observations and if some dust models are incompatible with the data. The RT modelling was
therefore repeated for a number of dust models.

Some of the dust models have been tuned to reproduce observations of the Galactic diffuse medium,
such as the Compiègne et al. model \citep[][COM below]{Compiegne2011} and the THEMIS 2.0 model
\citep[see Table 3 in][]{Ysard2024}. Other models are more representative of dense regions in which
the grains are thought to have evolved by coagulation and possibly by accretion of ice mantles. We
use three types of model for these evolved cases.

The first type is described in \citet{Kohler2015} and is based on the optical properties of THEMIS 1.0
\citep{Jones2013}\footnote{The main difference between THEMIS 1.0 \citep{Jones2013} and THEMIS 2.0
\citep{Ysard2024} is the choice of optical properties for silicates. The first version of the model
uses the properties described in \citet{ScottDuley1996} down to the mid-IR and an extrapolation of
those into the FIR. The second version of the model uses the optical properties of silicates
measured in the laboratory by \citet{Demyk2017, Demyk2022} down to $\lambda \sim 1$ mm.}. It
includes three types of large grains: CMM (core/mantle/mantle grains),  where nano-grains are
considered to have accreted onto the surface of the big carbons and silicates; their aggregated
forms, AMM; and these same aggregates with an additional water ice mantle (AMMI). A summary of the
properties of these grains is given in Sect. 2.3 of \citet{Jones2017}.

The second type of model is described in \citet{Ysard2019} and is still based on THEMIS 1.0
\citep{Jones2013}. Three types of large grains are considered: (i) Mix 1, which is large spherical
grains composed, by volume, of 2/3 silicate and 1/3 carbon; (ii) Mix 1:50, which is the same as the
previous ones but with a porosity of 50\%; (iii) Mix 1:Ice, which is the same compact grains as Mix
1, to which a mantle of water ice representing 55\% of their volume has been added.

The third type of model is built on the same principle as Mix 1(:50:Ice) coarse grains but using
the optical properties of THEMIS 2.0 \citep{Ysard2024}, not THEMIS 1.0. Four types of coarse grains
are considered: (i) mixed-mantle grains (MC), in which the grains are a mixture of carbon and
silicate; (ii) mixed-core porous (MCP), which are the same as MC but with a porosity of 25\%; (iii)
mixed-core ice (MCI), MC to which a mantle of water ice representing 50\% of their volume has been
added; (iv) MCPI, MC to which a mantle of water ice representing 50\% of their volume has been
added. Appendix~\ref{app:single} shows further examples of opacities and scattering functions (as
traced by the asymmetry parameter $g$) for these dust models.

\begin{figure}
\begin{center}
  \includegraphics[width=9cm]{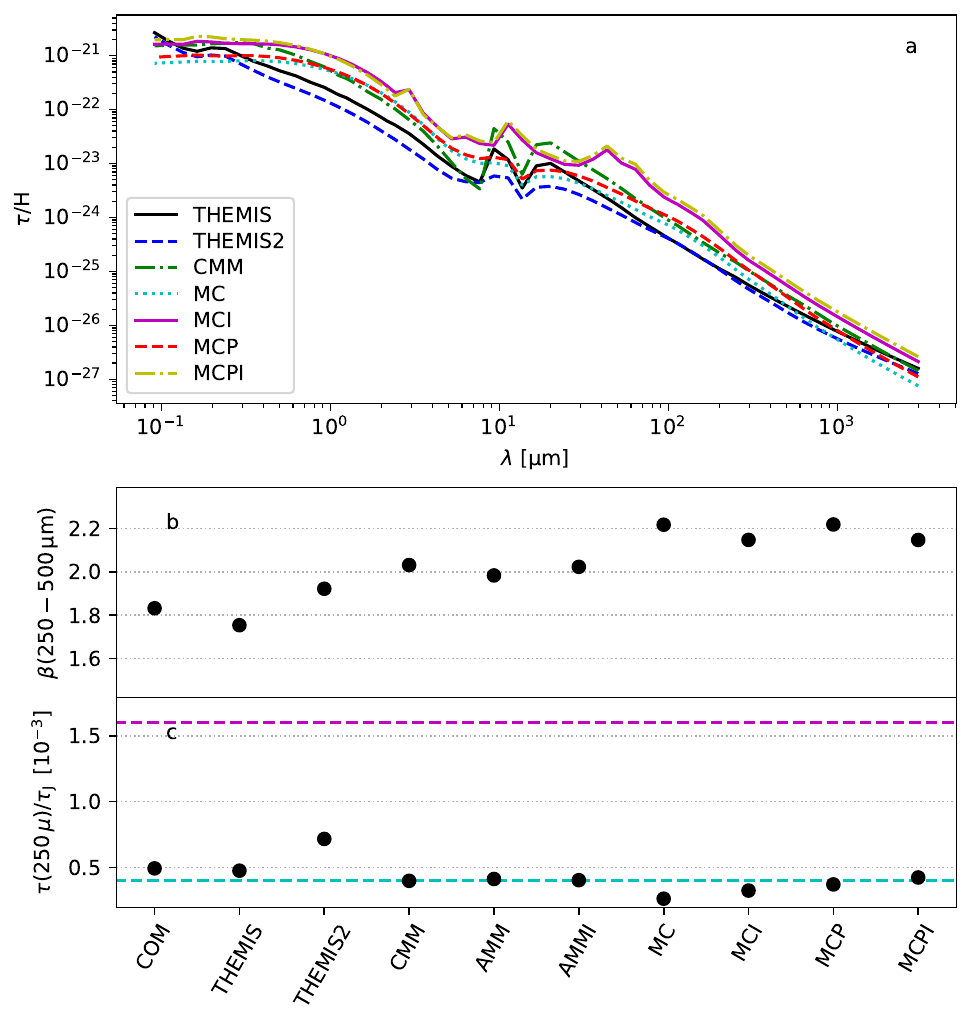}
\end{center}
\caption{
Extinction curves for selected dust models (frame a). Frame b shows the values of the FIR opacity
spectral index. Frame c shows the FIR-NIR optical depth ratios $\tau(250\,\mu{\rm m})/\tau({\rm
  J})$, where the cyan dashed line is drawn at $\tau(250\,\mu{\rm m})/\tau({\rm J})=0.4 \times
10^{-3}$ and the magenta dashed line at $\tau(250\,\mu{\rm m})/\tau({\rm J})=1.6 \times
10^{-3}$. The THEMIS2 models (MC to MCPI) have their default grain size distributions with $a_0\sim
0.05\,\mu$m.
}
\label{fig:plot_dusts_A}
\end{figure}

\begin{figure}
\begin{center}
\includegraphics[width=9cm]{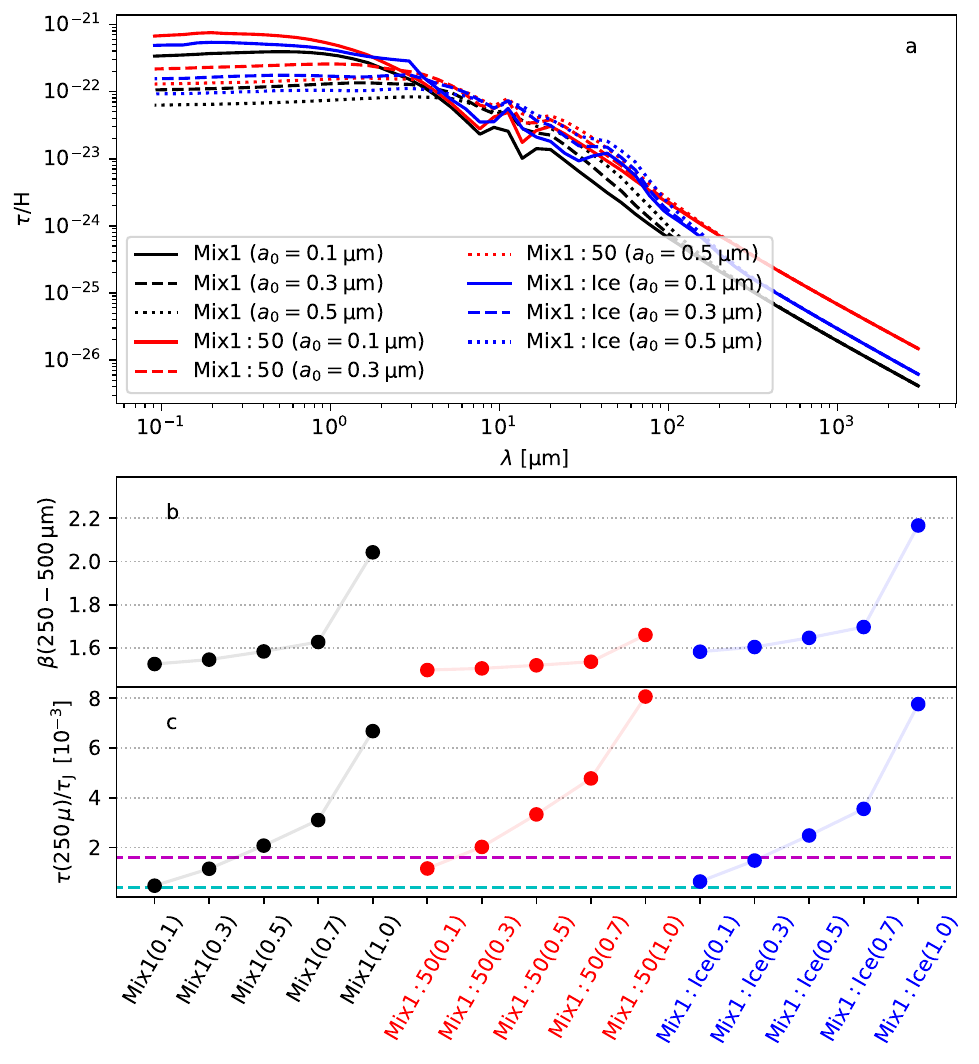}
\end{center}
\caption{
As Fig.~\ref{fig:plot_dusts_A} but for  THEMIS dust models Mix 1, Mix 1:50, and Mix 1:Ice. Data are
shown for size distributions with $a_0=$0.1, 0.3, 0.5, 0.7, and 1.0 $\mu$m (numbers quoted within
parentheses).
}
\label{fig:plot_dusts_B}
\end{figure}

Figures~\ref{fig:plot_dusts_A}-\ref{fig:plot_dusts_B} show examples of dust extinction curves
and the 250-500\,$\mu$m dust opacity spectral index $\beta$.  The $\beta$ values are mainly in the
range of $\beta\sim 1.5$ to $\beta \sim$2.2.  Through their effect on dust temperatures, they
also have an effect on the fitted ISRF levels. 

One key dust parameter is the opacity ratio between the short wavelengths, where dust absorbs
energy, and the long wavelengths, where the energy is again re-emitted away. In the present study,
the NIR observations directly constrain the absolute level of extinction, while the quality of the
FIR fits is mainly affected by the shape of the FIR extinction curve.
Figures~\ref{fig:plot_dusts_A}-\ref{fig:plot_dusts_B} use $\tau(250\,\mu{\rm m})/\tau({\rm J})$ as
a proxy for the FIR-to-NIR ratio. The ratio has been estimated to be $\tau(250\,\mu{\rm
  m})/\tau({\rm J}) \sim 0.4 \times 10^{-3}$ or slightly higher in diffuse clouds
\citep[e.g.][]{Planck2014_allsky_model}, but the ratio tends to be higher in molecular
clouds. \cite{GCC-V} found a four times higher median ratio, $\tau(250\,\mu{\rm m})/\tau({\rm
  J}) \sim 1.6 \times 10^{-3}$, for a sample of dense clumps observed with {\it Herschel}. Most
plotted dust models have $\tau(250\,\mu{\rm m})/\tau({\rm J})$ ratios that are more consistent with
observations of low-density clouds, but the ratio $\tau(250\,\mu{\rm m})/\tau({\rm J})$ increases
rapidly with increasing grain sizes. A complete list of the $\beta$ and $\tau(250\,\mu{\rm
  m})/\tau({\rm J})$ values of the dust models tested is included in Appendix~\ref{sect:dust_table}.

The dust properties were initially kept spatially constant, but later we also tested models with a
smooth transition from one set of dust properties at low densities to another at higher
densities. The transition was implemented by adjusting the relative abundances of two dust models;
the relative abundances [0,1] of the first dust are
\begin{equation}
\chi = \frac{1}{2} - \frac{1}{2} \tanh [ \xi \times (\log_{\rm 10} n_{\rm H}-\log_{\rm 10} n_{\rm
    0} )].
\label{eq:abu}
\end{equation}
The second dust has abundances $1-\chi$ and thus becomes dominant at high volume densities. The
density threshold of the transition is given by $n_{\rm 0}$ and the parameter $\xi$ , which determines the
steepness of the transition and was set to $\xi=4$.

\subsection{Model fits}  \label{sect:fits}

Radiative transfer models were optimised to find the best match to the 250-500\,$\mu$m FIR
data. The free parameters in the fits are one scalar factor $k_{\rm ISRF}$ for the radiation field
intensity and one scaling factor $k_{\rm N}$ for the column density associated with each map pixel.

The factor $k_{\rm ISRF}$ was optimised based on the ratios of the observed and model-predicted
$I_{\nu}(250\,\mu{\rm m})/I_{\nu}(500\,\mu{\rm m})$ ratios that were averaged over the $Q=1-2$
regions (Fig.~\ref{fig:make_masks}). The optimisation thus concentrated on the dense parts of the
field, also because the observations of low column density areas are more likely to be affected by
large-scale density and radiation-field variations outside the modelled volume. The $k_{\rm N}$
values were updated based on the ratio of the observed and model-predicted 350\,$\mu$m surface
brightness, and therefore the models have a nominal resolution of 30$\arcsec$.

The combination of LOS cloud shapes, $A^{\rm ext}_{\rm V}$ values, and dust models resulted in
a large number of models (cf. Table~\ref{table:dust_table}). The quality of each fit was quantified
using the relative error
\begin{equation}
r = 100\% \times \frac{I_{\nu}^{\rm MOD}-I_{\nu}^{OBS}}{I_{\nu}^{OBS}},
\end{equation}
where $I_{\nu}^{\rm OBS}$ and $I_{\nu}^{MOD}$ are the observed and model-predicted surface
brightness maps, respectively.
The comparison was done at 30$\arcsec$ resolution, except at 500\,$\mu$m where the model maps
were convolved down to the 41$\arcsec$ resolution of the corresponding {\it Herschel} map.
The quantity $b=\langle r \rangle$ is the bias and $\sigma_r$ the standard deviation of the $r$
values. These can be evaluated separately for a given wavelength and for different regions
(different $Q$ values; Sect.~\ref{sect:masks}).  We also used the mean squared error (MSE). While
$\sigma$ measures variation relative to the mean residual, MSE measures the variation relative to
zero and is thus also affected by systematic errors (bias). The model optimisation was stopped when
the 350\,$\mu$m maps indicated good convergence with $b(350\,\mu{\rm m})<0.3$\,\% and
$\sigma(r(350\,\mu{\rm m}))<1$\,\%. The change in the radiation field intensity per iteration was
also required to be small, with $\Delta k_{\rm ISRF}<0.2\%$.  One typically reached $\sigma(r)\sim
0.5$\%, except for a subset of models where the residuals in the densest filaments remained
positive due to the saturation of the surface brightness (i.e. the model surface brightness
remaining below the observed values).

The model predictions were also compared to the observations of the 160\,$\mu$m surface brightness
and the NIR extinction. These data, however, were not used in the optimisation itself.

\section{Results} \label{sect:results}

\subsection{Models with constant dust properties} \label{results:one}

\begin{figure}
\begin{center}
\includegraphics[width=9cm]{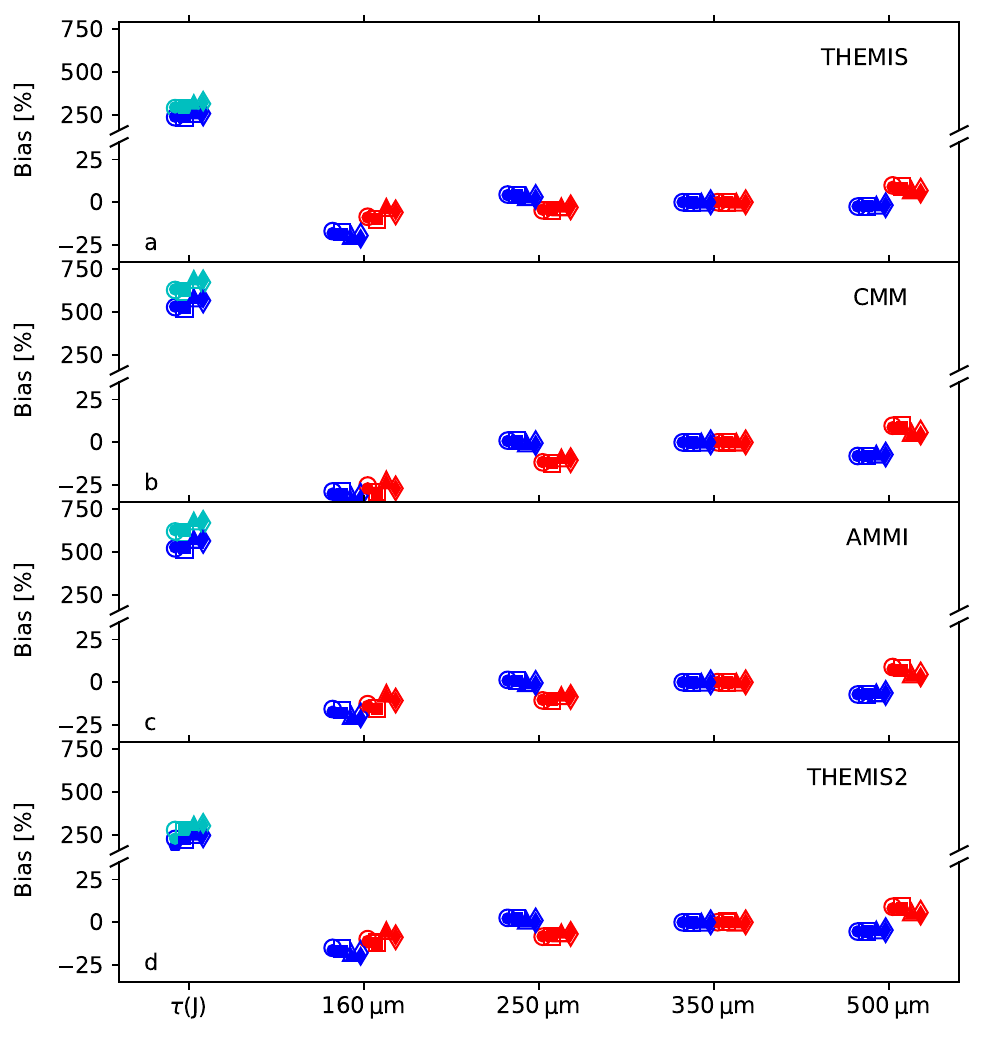}
\end{center}
\caption{
Bias $\langle (I_{\nu}^{\rm MOD} - I_{\nu}^{\rm OBS})/ I_{\nu}^{\rm OBS} \rangle$ for four dust
models (frames a-d). The symbols correspond to the cloud models A-D (circles, squares, triangles,
and diamonds, respectively) that show here only minor differences.  The x-axis label $\tau({\rm
  J})$ refers to the filament ($Q=1$) J-band opacity, where the models are compared to the
extinction derived with background stars and the \citet{Indebetouw2005} extinction curve, using
either the NICER map (blue symbols) or median value of individual stars over the selected
  region (cyan symbols). For 160-500\,$\mu$m the errors are shown for the filament ($Q=1$, blue
symbols) and the ridge ($Q=2$, red symbols) regions, assuming an external extinction of $A^{\rm
  ext}_{\rm V}=0.5^{\rm m}$ (open symbols) or $A^{\rm ext}_{\rm V}=1.0^{\rm m}$ (filled
symbols). The symbols are spread along the x-axis for better readability.
} 
\label{fig:plot_single_1}
\end{figure}

We first examined models with spatially constant dust properties.  Figure~\ref{fig:plot_single_1}
shows the fit bias in the case of four dust models, using four cloud density fields (A-D;
cf. Table~\ref{table:clouds}), and for the values $A^{\rm ext}_{\rm V}=0.5^{\rm m}$ and $1^{\rm m}$
of the external attenuation. The results, and especially the $\tau({\rm J})$ bias, are seen to depend
mainly on the dust model, with smaller variations due to cloud shape and $A^{\rm ext}_{\rm V}$
values.  The models of Fig.~\ref{fig:plot_single_1} tend to slightly underestimate the 160\,$\mu$m
intensity, while the NIR opacity $\tau({\rm J})$ is more significantly overestimated.

\begin{figure}
\begin{center}
\includegraphics[width=9cm]{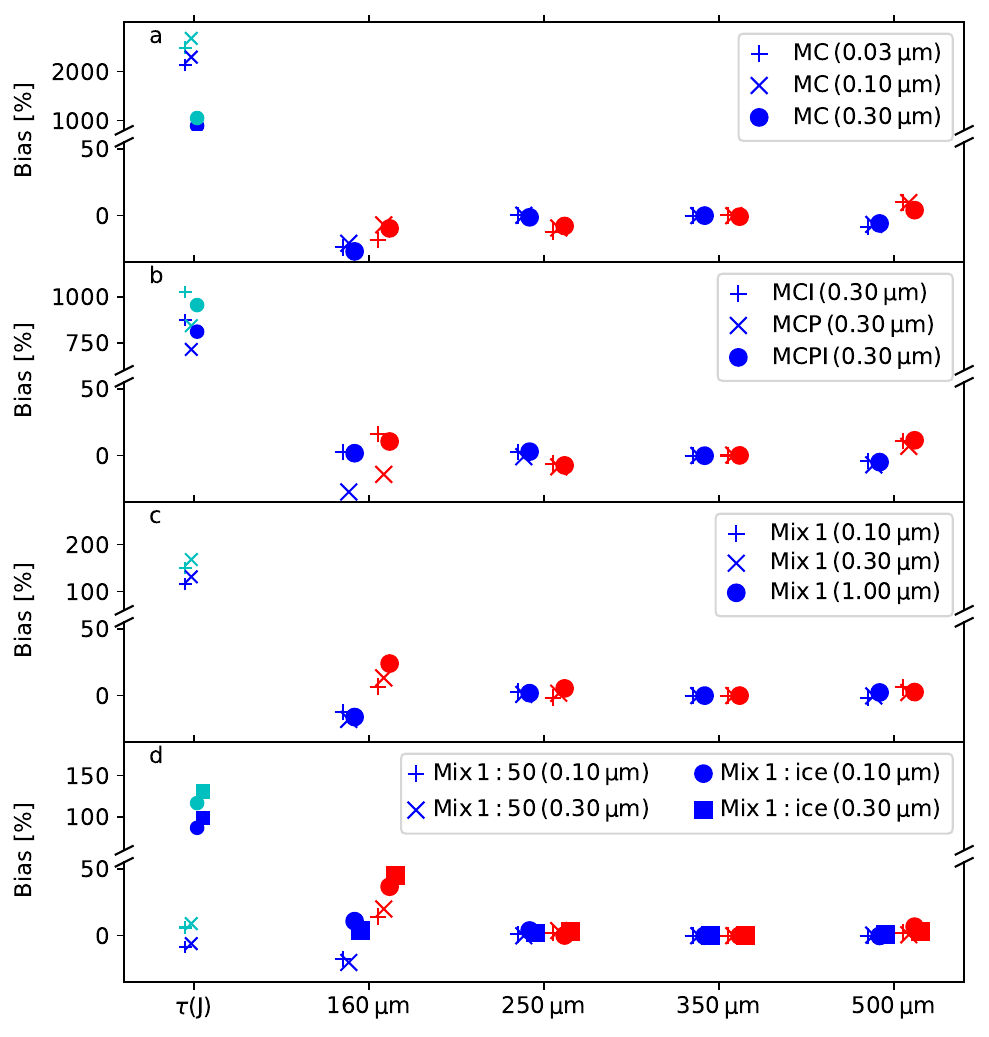}
\end{center}
\caption{
As Fig.~\ref{fig:plot_single_1}, but showing results for larger sample of dust models. The cloud
model is A with $A^{\rm ext}_{\rm V}=1$\,mag.
} \label{fig:plot_single_2}
\end{figure}

Figure~\ref{fig:plot_single_2} shows results for a set of 13 dust models. The figure also probes
the effects of different grain size distributions (e.g. $a_0$=0.1\,$\mu$m vs. $a_0$=0.3\,$\mu$m),
but regarding the NIR opacity, the results are still affected more by dust type than by size
changes. The Mix 1 models of frames c and d are in good agreement with the $\tau({\rm J})$
measurements and with only moderate positive bias for the extrapolated 160$\mu$m surface
brightness, mainly in the ridge region.

\begin{figure*}
\sidecaption
\includegraphics[width=11.8cm]{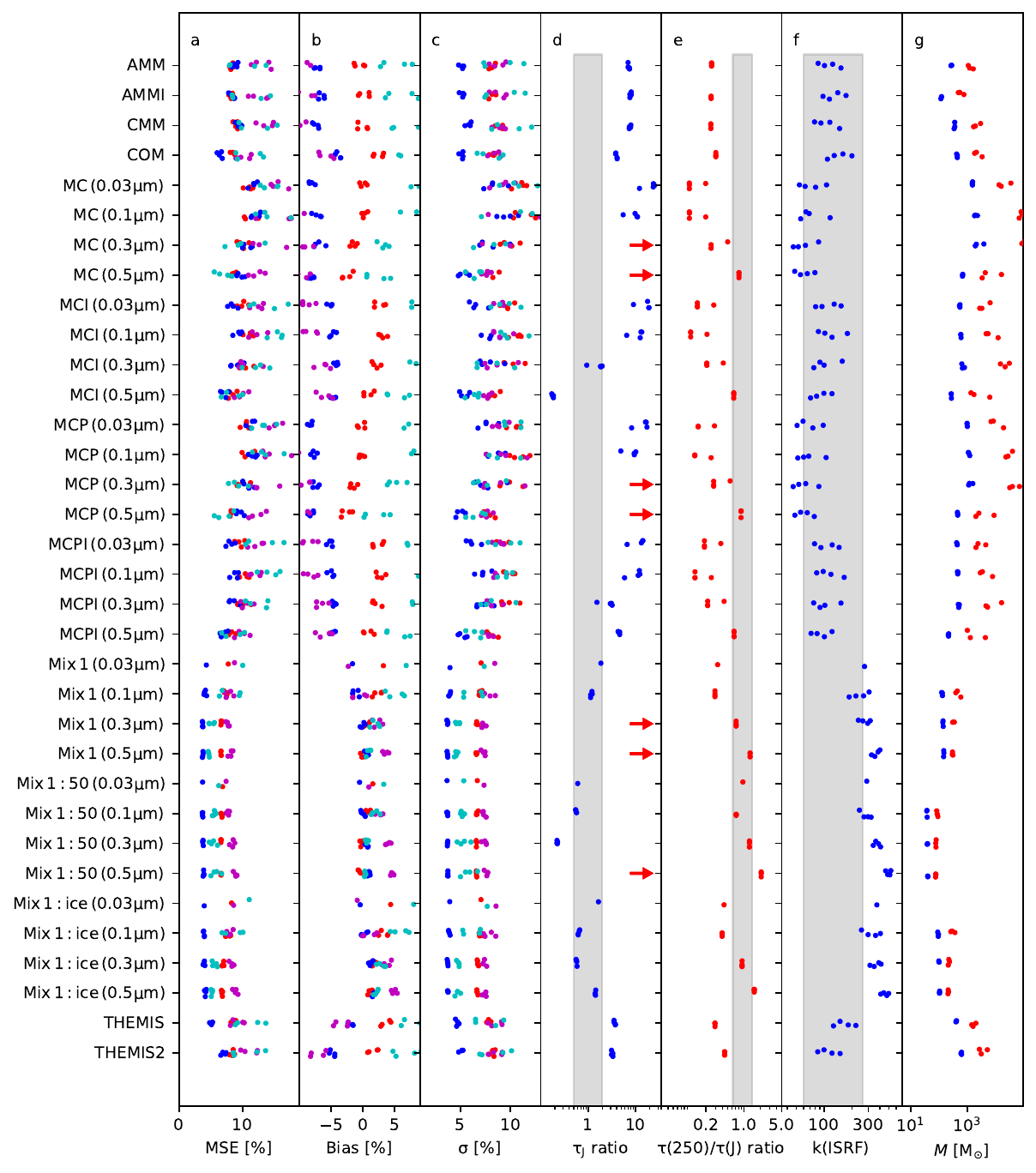}
\caption{
Results for single-dust cases for cloud models $A$-$D$ and  $A^{\rm ext}_{\rm V}$=1\,mag. Frames
a-c show the bias, the standard deviation $\sigma$ of the residuals, and the MSE
  values. These are shown for 250\,$\mu$m (red and magenta dots for $Q=1$ and $Q=2$,
respectively) and for 500\,$\mu$m (blue and cyan dots for $Q=1$ and $Q=2$, respectively).
Frame d shows the ratio of the J-band extinction predicted by the cloud model and the median
estimate, based on background stars and the extinction curve of the dust model in question. The
shading corresponds to a factor of two uncertainty, and the red arrows indicate values outside the
plotted range.
Frame e shows the mean ratio $\tau(250\mu{\rm m})/\tau({\rm J})$ in the $Q=1$ area, relative to the
reference value $\tau(250\mu{\rm m})/\tau({\rm J})=1.6\times 10^{-3}$ (red points; with $\tau({\rm
  J})$ for the standard extinction curve), the shading corresponding to $1-2.3\times 10^{-3}$
\citep{GCC-V}.
Frame f shows the $k_{\rm ISRF}$ values for $A^{\rm ext}_{\rm V}=1$\,mag fits, the shading
indicating the range 60-260.
The masses of the model clouds are shown in Frame g (blue for the $Q=1$ and red for the $Q=2$
area).
}
\label{fig:fit_plots_3}
\end{figure*}

Figure~\ref{fig:fit_plots_3} shows the fit quality, estimated radiation field intensity, and
filament mass for selected models. The FIR residuals are plotted separately for the filament and
ridge regions ($Q=1$ and $Q=2$, respectively). The model-predicted $\tau({\rm J})$ values are
compared with the average VISION extinction value over the $Q=1$ area, where the mean value over the
$Q=-1$ area is subtracted as the background. Each line on the y-axis corresponds to one dust model,
including all cloud models A-D and $A^{\rm ext}_{\rm V}=1^{\rm m}$.
There is noticeable variation in fit quality. The MSE values range from $\sim$5\% up to 10\%
and even 20\% in the ridge region (Fig.~\ref{fig:fit_plots_3}a). The bias ranges from close to zero
to $\pm$5\% and even beyond (Fig.~\ref{fig:fit_plots_3}b). The $\sigma$ values are typically
5-10\% and tend to be higher at 250\,$\mu$m than at 500\,$\mu$m (Fig.~\ref{fig:fit_plots_3}c).

Frame d shows the ratio between the J-band optical depth in the $Q=1$ area in the model and as
estimated from the reddening of the background stars, using the NICER method and the extinction
curve of the dust model in question. In most cases the NIR extinction is overestimated, and
  especially models with $a_0\ge 0.5\,\mu$m appear to be clearly excluded.

If $\tau(250\,\mu{\rm m})/\tau({\rm J})=1.6\times 10^{-3}$ were assumed to be the correct value
also for OMC-3, then models with large grains, such as Mix 1, Mix 1:50, and Mix 1:Ice with
$a_0=0.3\,\mu$m would be preferred (Fig.~\ref{fig:fit_plots_3}e). However, in \citet{GCC-VI} this
ratio was derived assuming the normal NIR extinction curve and is therefore internally inconsistent in
the case of large grain sizes. If large grains $a_0 \ga 0.3\,\mu$m were present everywhere in the
OMC-3 field, the NIR extinction curve would be flat (cf. Fig.~\ref{fig:plot_dusts_B}), and the
NICER map should show only little correlation with the column density derived from dust
emission. Based on Fig.~\ref{fig:plot_extinction}, this loss of correlation may be observed in the
ridge region. However, the ridge is probed by only a small number of background stars, and this
prevents strong conclusions. Large grain sizes would also result in higher estimates for the
filament mass and especially the radiation field intensity (Fig.~\ref{fig:fit_plots_3}f-g).

Figure~\ref{fig:fit_plots_3} does not include models that failed the convergence criteria.  Of the
more than 500 models examined, only 23 showed greater than 5\% positive 350\,$\mu$m residuals in the
ridge area. This indicates saturation of the model-predicted signal, the surface brightness no
  longer increasing with increasing column density. These cases corresponded mainly to the dust models MC, MCI,
  MCP, or MCPI  with $a_0=0.1$ or $0.3\,\mu$m, often combined with a harder radiation
  field with $A_{\rm V}^{\rm ext}=-0.5^{\rm m}$.  

The approaching saturation of surface brightness can increase the mass ratio between the ridge and
filament regions, even for some models included in Fig.~\ref{fig:fit_plots_3}.  This is not likely
to be due to internal heating. First, the masses were estimated for areas where the MIR and FIR
point sources were already masked. Second, when we tested models with internal heating sources
(Sect.~\ref{sect:radiation}) and optimised their luminosities based on their local impact on
surface brightness, the effect on the total mass in the $Q=1$ and $Q=2$ regions remained always
insignificant.

Figure~\ref{fig:plot_single_3} in Appendix~\ref{app:single} shows selected data for 80 models with
the lowest MSE values (combining the 250\,$\mu$m and 500\,$\mu$m errors in the $Q=1-2$
regions). These low-MSE models employ different versions of the Mix 1, Mix 1:Ice, and Mix 1:50 dust
models with grain size distributions $a_{\rm 0}=0.1-0.7\,\mu$m. In this sample, the MSE values vary
only little, but the match to the observed $A_{\rm J}$ values is sensitive to the grain size.
  At 160\,$\mu$m the models often overestimate the filament ($Q=1$) surface brightness by
  nearly 30\%, while in the ridge region ($Q=2$) the residuals are scattered more around zero.
  Several of the $a_{\rm 0}=0.1\,\mu$m dust models match the observed NIR extinction to within
  30\%. However, cases $a_{\rm 0} \ge 0.3\,\mu$m tend to result in large positive or negative
  extinction estimates, which would exclude the presence of such large grains over the
  whole field.

There is no strong preference between different LOS density profiles. All A-D cloud models appear
among the best-fitting models. The values of external extinction $A^{\rm ext}_{\rm V}$=0.5\,mag and
$A^{\rm ext}_{\rm V}$=1\,mag are also equally common. There are even a few models with $A^{\rm
  ext}_{\rm V}=-0.5$\,mag, the case with the harder radiation field. The fit quality is relatively
insensitive to the assumed spectral shape of the radiation field because the absolute level of the
radiation field was always adjusted as a separate free parameter.

\subsection{Models with spatial dust variations} \label{results:two}

Spatial dust property variations were examined using combinations of two dust models whose
relative abundances were varied based on the volume density (Sect.~\ref{sect:dust}). Because
previous tests showed only little dependence on the other model parameters, the tests were done
using only the cloud model $C$ with $A_{\rm V}^{\rm ext}=1$\,mag.

\begin{figure*}
\sidecaption
\includegraphics[width=12cm]{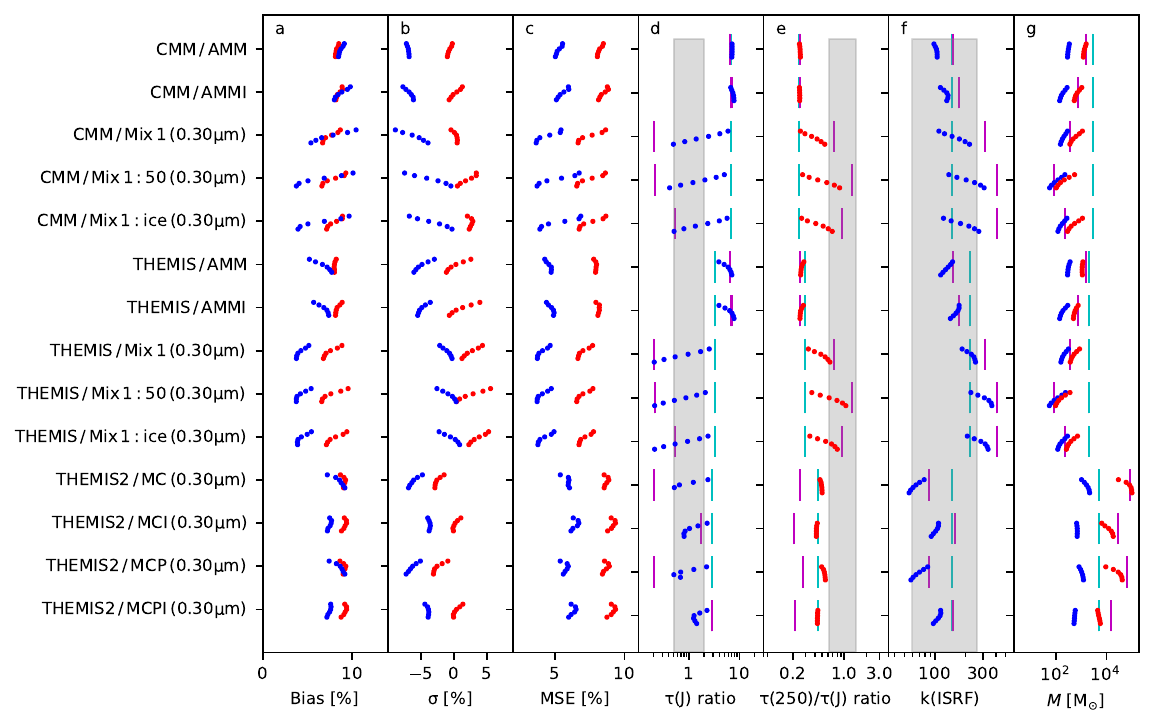}
\caption{
As Fig.~\ref{fig:fit_plots_3}, but for models with two dust components with density-dependent
abundances. Results are shown for the cloud model C with $A_{\rm V}^{\rm ext}=1$\,mag and density
thresholds $n_0=10^{3}-3.2\times 10^{4}\,{\rm cm}^{-3}$, with symbols for higher $n_0$ values
shifted to slightly higher y-axis positions. In frame d, $\tau({\rm J})$ is calculated using the
actual dust models and the spatially varying extinction curves. The vertical lines in frames d-e
indicate the corresponding single-dust results from Fig.~\ref{fig:fit_plots_3} for the first (cyan
lines) and the second (magenta lines) of the dust components.
}
\label{fig:fit_plots_3_two}
\end{figure*}

Figure~\ref{fig:fit_plots_3_two} shows the results for the examined dust combinations. The density
threshold $n_0$ for the transition between the two dust models was varied from $10^3$ to $3.2\times
10^4$\,cm$^{-3}$ in steps of a factor of two. Thus, for the highest (lowest) threshold, the results are
similar to the single-dust models that used the same dust model as now in the low-density
(high-density) region.

The combination of THEMIS dust at low densities and some Mix dust at high densities
results in marginally the lowest bias values and the lowest dispersion $\sigma$ in the
residuals. However, the fits get consistently better with lower $n_0$, approaching the
Fig.~\ref{fig:fit_plots_3} results where the same Mix dust is used for the whole
model. The lower $n_0$ requires in this case a higher radiation field intensity, up to $k_{\rm
  ISRF}\sim 400$.  The same is observed for the combination of CMM and Mix 1:50 or Mix
1:Ice, where the best fits are obtained for low values of $n_0$ that again correspond to
high values of $k_{\rm ISRF}$.

The combinations of THEMIS2 at low densities and MC, MCI, MCP, or MCPI dust at high densities have
relatively low $\sigma$ values. They provide a good fit to the observed NIR extinction, but also show
higher overall MSE values.
Their $\tau({\rm 250\,\mu m})/\tau({\rm J})$ ratios are slightly below the range of values
  that \cite{GCC-V} found for dense clumps (the shaded area in
  Fig.~~\ref{fig:fit_plots_3_two}e). However, the OMC-3 field is also quite different from that
  previous sample, which consisted mainly of clumps in much less active environments.
What is not visible in
Fig.~\ref{fig:fit_plots_3_two} is that for $n_0\la 8000$\,cm$^{-3}$ the modelled surface brightness
saturates for these dust combinations, resulting in significant positive residuals along the
ridge. This was observed less in the case of single-dust models and is thus acerbated by stronger
attenuation of the radiation field in the outer parts of the cloud (due to the THEMIS2 dust) and
the resulting low temperature and weak emission of the MC dust along the ridge. In this case, the
single-dust models perform better than their combination.

The tested dust combinations provide roughly equally good fits to the FIR data, and the main
discriminator is the NIR extinction. However, the extinction can also be matched with almost pure
Mix 1, Mix 1:50, or Mix 1:Ice dust, which also have $\tau({\rm 250\,\mu m})/\tau({\rm J})$ ratios
  similar to those found in other dense clumps \citep{GCC-V}.

\section{Discussion} \label{sect:discussion}

We used RT modelling to investigate dust emission in the OMC-3 region. The
starting points were the {\it Herschel} FIR observations of dust emission, but also the model
predictions for the NIR extinction were later compared to the observations.

The FIR emission maps could be reproduced by almost all of the dust models tested, with only small
variations in fit quality. The dust models mainly varied in how they matched the outer filament
($Q$=1) relative to the ridge ($Q$=2). The largest fit residuals were typically seen towards the
high-density ridge, where the surface brightness is close to the maximum that some dust models can
produce before saturation. Actual saturation was observed, especially when the THEMIS2 dust model in
the outer cloud regions was combined with MC, MCI, MCP, or MCPI dust in the high-density part.  An
abrupt change in model dust properties could also result in unphysical filament profiles with a
sharp jump in the density. In our models this was not clearly visible, partly due to the gradual
change in the density-dependent relative abundances of the dust components and partly because the
contributions from the material below and above $n_0$ change smoothly in the 2D projection.

One caveat is that our calculations do not include the thermal coupling between gas and dust, which
could be important at high densities if the gas temperature without the coupling were much
higher. \citet{Goldsmith2001} estimated that at densities $n({\rm H}_2) \sim 10^6\,{\rm cm}^{-3}$
the dust temperature could increase by more than one degree. The result corresponded to the
normal ISRF attenuated by a factor of $10^{-4}$. In OMC-3 the radiation field could be two orders
of magnitude stronger, but this is also compensated for by higher extinction. The visual
  extinction reaches $A_{\rm V}\sim 10$\,mag at $\sim 1.5\arcmin$ distance from the filament centre
  (attenuation by a factor of $\sim e^{10/2.5} \approx 55$). At that point, most UV photons have
  already been absorbed, thus eliminating most of the photoelectric heating. It therefore seems 
unlikely that the gas-to-dust coupling would cause noticeable effects within the central filament.

Several parts of the ridge ($Q=2$) were masked due to resolved MIR point sources or very strong FIR
emission. These areas were excluded when the intensity of the external field was optimised.  The
additional heating from young stars and partly unresolved protostellar sources could be important
in those areas and might explain the failure of some models to converge due to positive
350\,$\mu$m residuals in the $Q\ge2$ areas (Table~\ref{table:dust_table}). However, when embedded
sources were included in models and the luminosities were optimised based on the local surface
brightness, the effect of the sources remained very localised. The effect of strong sources had 
already been eliminated by direct masking ($Q>2$), and their impact on the wider filament region ($Q=1$)
should be low.

Figure~\ref{fig:plot_examples_1} in Appendix~\ref{app:single} shows an example of the fit
residuals.  Embedded sources are not part of that model, and any local heating should result in
compact positive residuals at short wavelengths. However, the errors show mainly extended structures
that are probably correlated with variations in the cloud structure and external heating that are
not perfectly matched by the model. The residual maps are smooth, especially at 500\,$\mu$m, where
the errors vary by no more than $\sim$10\% across the filaments.

Because the column density updates were based on 350\,$\mu$m data, the 250\,$\mu$m and
500\,$\mu$m residuals are anticorrelated. The 160\,$\mu$m residuals tend to have a spatial
distribution similar to the 250\,$\mu$m residuals,  but due to the greater distance from the 350\,$\mu$m
reference wavelength, a larger magnitude. The shorter wavelengths are also more sensitive to
variations in the radiation field and in the diffuse outer cloud layers that may be partly outside
the modelled volume. In Fig.~\ref{fig:plot_examples_1}, the 160\,$\mu$m residuals are, on average,
close to zero towards the centre of the map (where $k_{\rm ISRF}$ was determined) but increase
strongly towards the map edges. Figure~\ref{fig:plot_single_3} shows that for the best single-dust
models the 160-250\,$\mu$m bias tends to be higher (less negative or more positive) in the ridge
($Q=2$) compared to the surrounding filament ($Q=1$).
The 160\,$\mu$m bias values are around -30\% in the filament. In the ridge, the 160\,$\mu$m
  bias is close to zero for the best models, with an overall range from -5\% to 30\%. The
  250\,$\mu$m bias is small in the filament, between -1\% and +2\%, but systematically positive in
  the ridge, with values up to $\sim$5\%.
Errors at the level of a few per cent could be attributed to uncertainties in the calibration and,
especially at 160\,$\mu$m, to uncertainty in the background subtraction (Sect.~\ref{sect:bgsub}),
but the larger errors are significant.
Models that give the correct 250\,$\mu$m intensities in the filament region also tend to match the
160\,$\mu$m observations of the ridge. In detail, the relative errors depend on the choice of 
grain materials and the shape of the size distribution.

\subsection{Effects of cloud shape and radiation field spectrum}

Modelling of low-mass star-forming clouds has shown clear dependence on the LOS density
distribution of the clouds, as well as some dependence on the spectrum of the illuminating radiation
field \citep{Juvela_L1642,Juvela2024_Taurus}. In contrast, in the OMC-3 models these were only
marginal factors, as shown by Figs.~\ref{fig:plot_single_1}-~\ref{fig:fit_plots_3}. This is likely
caused by the higher column density of the OMC-3 filaments, which results in almost all of the
incoming radiation being already absorbed in outer cloud layers. Although the observed intensity
does depend on the total column density, the FIR radiation observed towards the filaments
originates in the first approximation in plane-parallel layers on the front and back cloud surfaces.
Low-mass clouds are more transparent to dust-heating radiation, and a volume element is more
likely to receive radiation from multiple directions, making the heating more dependent on the
overall cloud shape.

Higher column density also decreases the dependence on the spectral shape of the illuminating
radiation field. For example, the peak of a $T_{\rm eff}$=38 000\,K star is in UV and a $T_{\rm
  eff}$=5000\,K black body peaks at $\sim 1\,\mu$m.  However, when the cloud has high column
density, UV and optical radiation is absorbed in the cloud surface layers. The rest of the cloud is
heated by the remaining NIR-MIR radiation, and there is no longer any difference between the
extincted $T_{\rm eff}$=38 000\,K and $T_{\rm eff}$=5 000\,K black bodies. This applies to the shape
of the radiation field, while the absolute level of the field is always a free parameter.

\subsection{Constraints from the radiation field intensity}

The models show a wide range of predictions for the radiation field intensity. This is partly due
to the fitted 250-500\,$\mu$m data covering only a narrow range of wavelengths. Since column
densities are also free parameters, a small change in the dust model $\beta$ can correspond to a
large change in the fitted $k_{\rm ISRF}$. The $k_{\rm ISRF}$ values also depend on the NIR-to-FIR
opacity ratios. If this ratio is large, the external radiation field is strongly attenuated, and
one can match the observed filament brightness only by increasing both its column density and the
absolute level of the external field.

The parameter $k_{\rm ISRF}$ corresponds to the radiation field at the outer boundary of the
models. We can make an independent estimate of the radiation field using background-subtracted
surface brightness measurements. Figure~\ref{fig:plot_relative_SED} shows selected NIR-mm spectral
energy distributions (SEDs) for the OMC-3 field, making use also of {\it Spitzer}, {\it WISE}, and
{\it AKARI} mid-infrared observations. If all the incoming radiation was absorbed and re-radiated,
the integral over the emission spectra should directly correspond to the $k_{\rm ISRF}$ values. The
SED in the brightest southern part of the filament corresponds to a radiation field with $k_{\rm
  ISRF}=262$. This is likely to be an upper limit, due to the presence of some local sources and
the surface brightness generally increasing towards the south. A position selected from
the filament wing gives $k_{\rm ISRF}=63$ and the average over the PACS area $k_{\rm
  ISRF}$=72. These are, in turn, likely to be lower limits if all of the incoming radiation is not
absorbed at the lower column densities. As a rough estimate, the expected values are $k_{\rm
  ISRF}$=100-200. In Fig.~\ref{fig:fit_plots_3}, some of the best dust models so far, Mix 1, Mix
1:50, and Mix 1:Ice, tend to require radiation fields above this range. On the other hand, most of
the other dust models are closer to the lower limit, $k_{\rm ISRF} \sim$100 or even below.

\subsection{Evidence of NIR extinction} \label{sect:extinction_curve}

NIR extinction estimates were calculated based on the reddening of the light of background
stars. The estimates depend on the assumed shape of the NIR extinction curve, and the variation
between different dust models is illustrated in Appendix~\ref{sect:dust_table}. The extinction
curves quoted in the literature correspond to only about 15\% variations in extinction, and some of the
tested dust models (e.g. THEMIS, THEMIS2, COM, CMM, and AMMI) give very similar values
(Fig.~\ref{fig:plot_cases}). However, the presence of larger grains $a_{\rm 0}\ga0.3\,\mu$m
flattens the NIR extinction curve and can result in a wide range of values and even some negative
extinction estimates. This affects the ratio of NIR opacities between the fitted model and the
NICER estimates. For example, models that use Mix 1:Ice dust with $a_0=0.3\,\mu$m have an NIR opacity
below the NICER estimates. The increase in size to $a_0=0.5\,\mu$m doubles the
$\tau(250\,\mu{\rm m})/\tau({\rm J})$ ratio, resulting in a significant reduction in column
density and the model NIR opacity. However, the flattening of the extinction curve reduces the
NICER estimates even more, so that the ratio between NIR model opacity and the NICER estimates
increases above one, instead of decreasing further (e.g. Fig.~\ref{fig:fit_plots_3} and
Fig.~\ref{fig:plot_single_3}).

Figure~\ref{fig:plot_extinction} shows that at larger scales (corresponding to $Q \le 1$) the FIR
filament is clearly visible in the \citet{Meingast2016} extinction map. This high degree of
correlation therefore implies an upper limit of $\sim 0.1-0.3$\,$\mu$m for the mean grain size in the
outer filament, the actual size limit depending on the dust model. In the ridge region ($Q=2$) the
correlation disappears, which could be an indication that the grains have reached sizes $\sim
0.3\,\mu$m. However, the strength of this evidence is reduced by the low number of background stars
detected in the $Q \ge 2$ area (Fig.~\ref{fig:make_masks}).

The size limits directly implied by the NIR data are roughly compatible with the evidence of the
fitted cloud models.  Figure~\ref{fig:fit_plots_3} shows that in the $Q=1$ region the NIR opacity
of the model clouds becomes inconsistent with the NIR stellar observations if the grain size is
allowed to increase to $a_0 \sim 0.5\,\mu$m. For some dust models, such as MCI, MIX 1, Mix 1:50, and
Mix 1:Ice, the preferred sizes are only slightly smaller, $a_0 \sim 0.1-0.3\,\mu$m.

The models with two dust components with spatially varying relative abundances attempted to
better match the expectation of the largest grain growth being limited to the regions of highest
volume density.  In the models of Fig.~\ref{fig:fit_plots_3_two}, the dust properties at low volume
densities are consistent with observations of the diffuse medium, and at high densities the grain size
is increased to $a_0=0.3\,\mu$m. The remaining free parameter was the density threshold $n_0$ of
the transition. The evidence from the goodness of fit is not very clear, as lower bias may be
associated with higher statistical errors, and the magnitudes of error in the $Q=1$ and $Q=2$
regions are often anticorrelated (Fig.~\ref{fig:fit_plots_3_two}a-c). However, with transition at
low densities, $n_0 \la 4000$\,cm$^{-3}$, the models generally were in better agreement with the NIR
extinction. The combination of two dust models naturally helps to retain the correlation between
the NIR extinction and the cloud structure seen in FIR data, while still allowing larger grains and
larger FIR emissivity in the central filament.

In principle, it would be useful to probe especially the $Q=2$ region using MIR point sources, to
investigate the potential changes in the NIR-MIR extinction curve. However, the {\it Spitzer} maps
of the ridge region contain only a handful of sources, mostly embedded young stellar objects rather
than background stars. The MIR extinction of the extended background surface brightness provides
additional constraints on dust properties but requires consideration of the additional
contribution of dust scattering and thermal emission \citep{Juvela2023_OMC3}. We will address the
extended MIR extinction in a future paper, in which the MIR emission from stochastically heated
grains will also be modelled.

The NIR scattered light could also be used to constrain the combination of grain properties and
the illuminating radiation field in the outer parts of the cloud, but accurate mapping of the
extended NIR surface brightness needs dedicated observations. Our models covered only the dense
part of the OMC-3 cloud, which is surrounded by column densities $N({\rm H_2})\sim
10^{21}$\,cm$^{-2}$ that also contribute to the NIR scattered surface brightness. Therefore, in
the present paper NIR scattering has only a small indirect role by increasing the photometric noise
and masking the faintest background stars.

\section{Conclusions} \label{sect:conclusions}

We presented modelling of the NIR-FIR dust observations of the OMC-3 region, concentrating on
its main filaments. The goal was to evaluate the possible range of dust properties and, as a
secondary point, to put constraints on the cloud structure and the radiation field intensity.  The
study led to the following conclusions:

\begin{itemize}
\item The 160-500\,$\mu$m observations of dust emission were always fitted satisfactorily, and this limited wavelength range alone is not sufficient to clearly
  reject any of the dust models.
\item NIR extinction measurements are a valuable additional constraint. Most dust models correspond
  to NIR opacities that are clearly higher than the direct extinction measurements based on
  background stars.
\item Large grain sizes result in the flattening of the NIR extinction curve. The good correlation
  between FIR emission and NIR extinction excludes the possibility of grain growth to mean sizes
  $a\sim 0.1\,\mu$m and above in the outer part of the filament. For the central ridge, the
  correlation is almost non-existent and the presence of large grains is possible.
\item The best overall fits to FIR observations were obtained with dust models Mix 1 and Mix 1:50 (large porous
  grains) with a mean grain size of $a_{\rm 0}=0.3\,\mu$m and with the Mix 1:Ice model with
  slightly larger grains with $a_{\rm 0}=0.5\,\mu$m. These have opacity ratios close to
  $\tau(250\,\mu{\rm m})/\tau({\rm J}) \sim 1.6 \times 10^{-3}$, the typical value previously seen
  in {\it Herschel} studies of dense clumps, when extinction is estimated using the 
  standard extinction curve.
\item Different dust models can result in very different predictions for the cloud mass
  and the local radiation field intensity. The RT runs with the otherwise
  best dust models, Mix 1, Mix 1:50, and Mix 1:Ice, require a slightly stronger radiation
  field in OMC-3 than the independently estimated 100-200 times the normal ISRF.
\item The fit quality of models with spatially varying dust properties was not clearly better than
  in the best single-dust models.

\item Overall, the combined FIR and NIR data are consistent with modest grain growth in the outer
  filament ($a\la 0.3\mu$m) and do not exclude further increases in size in the central ridge.
\end{itemize}
In the present paper we concentrate on the properties of large dust grains
that are in equilibrium with the radiation field and dominate the FIR
emission. In a future study, we will expand the investigation to dust emission
and scattering at MIR wavelengths to characterise the abundance and
emission properties of small grains.

\begin{acknowledgements}

MJ acknowledges the support of the Research Council of Finland Grant No. 348342.

\end{acknowledgements}

\bibliography{my.bib}

\begin{appendix}

\section{OMC-3 dust spectra}  \label{app:radiation_field}

\begin{figure}
\sidecaption
\includegraphics[width=8.8cm]{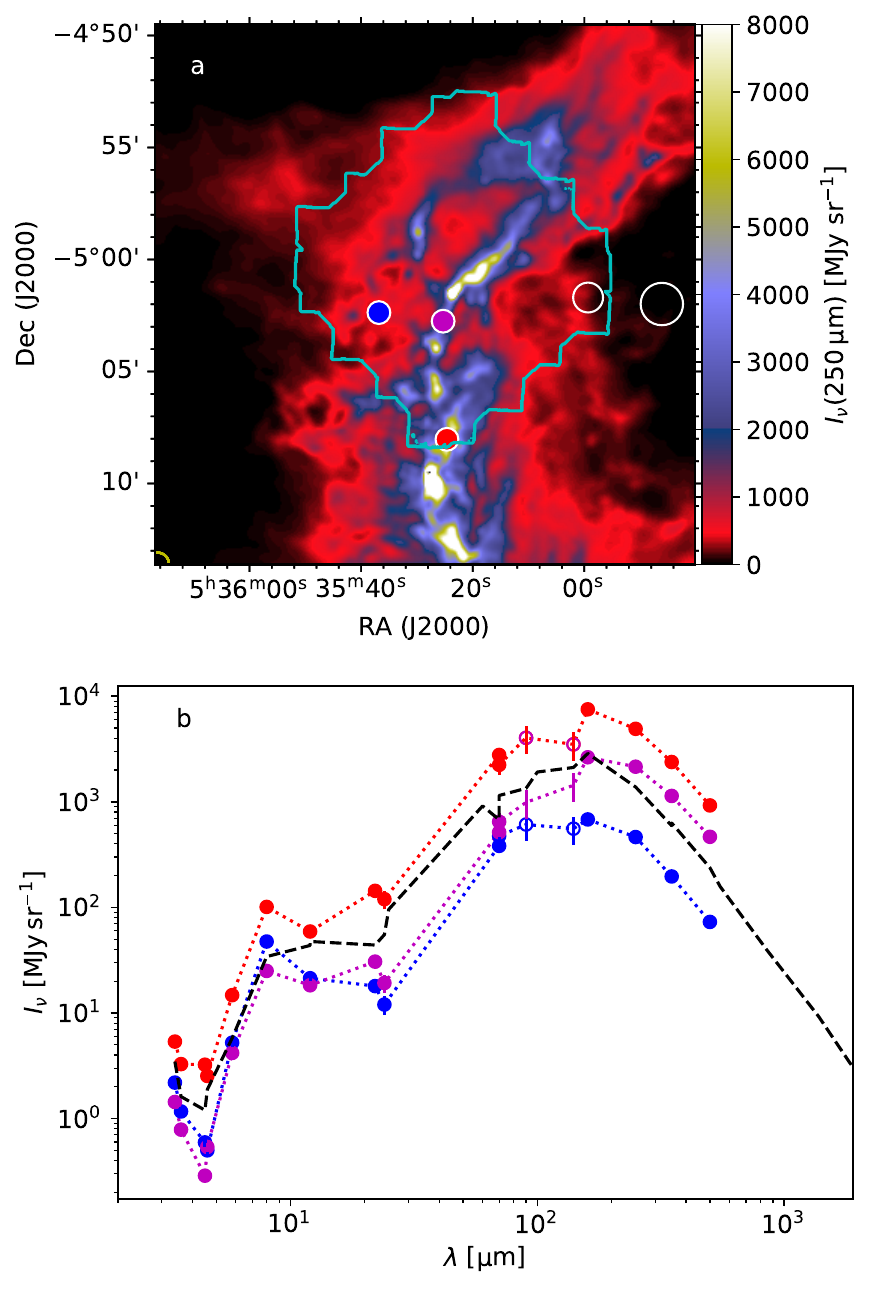}
\caption{
Sample spectra in OMC-3 field. Frame a shows the \Herschel 250\,$\mu$m map and the outline of
the PACS coverage. The white circles indicate the reference area used for background subtraction.
Frame b shows spectra for background-subtracted data at the positions marked in frame a with 
  blue, magenta, and red circles. The data correspond to 1$\arcmin$ (solid symbols; {\it Spitzer},
{\it WISE}, {\it Herschel}) or 2$\arcmin$ resolution (open symbols; {\it AKARI}, with 30\%
error bars). The average spectrum over the PACS coverage is drawn with a black dashed line.
}
\label{fig:plot_relative_SED}
\end{figure}

Figure~\ref{fig:plot_relative_SED} shows examples of the SEDs using background-subtracted
observations (Sect.~\ref{sect:bgsub}), giving an idea of the general SED shape and its
variations. Despite the high radiation field in the nearby Orion regions (both north and south of
OMC-3), the dust emission peaks beyond 100\,$\mu$m and thus corresponds to cool dust of $\sim$20\,K
for a wide range of dust models \citep[cf.][their Fig. 2]{Schuller2021}.

\section{Additional plots on single-dust models} \label{app:single}

\begin{figure}
  \sidecaption
  \includegraphics[width=8.8cm]{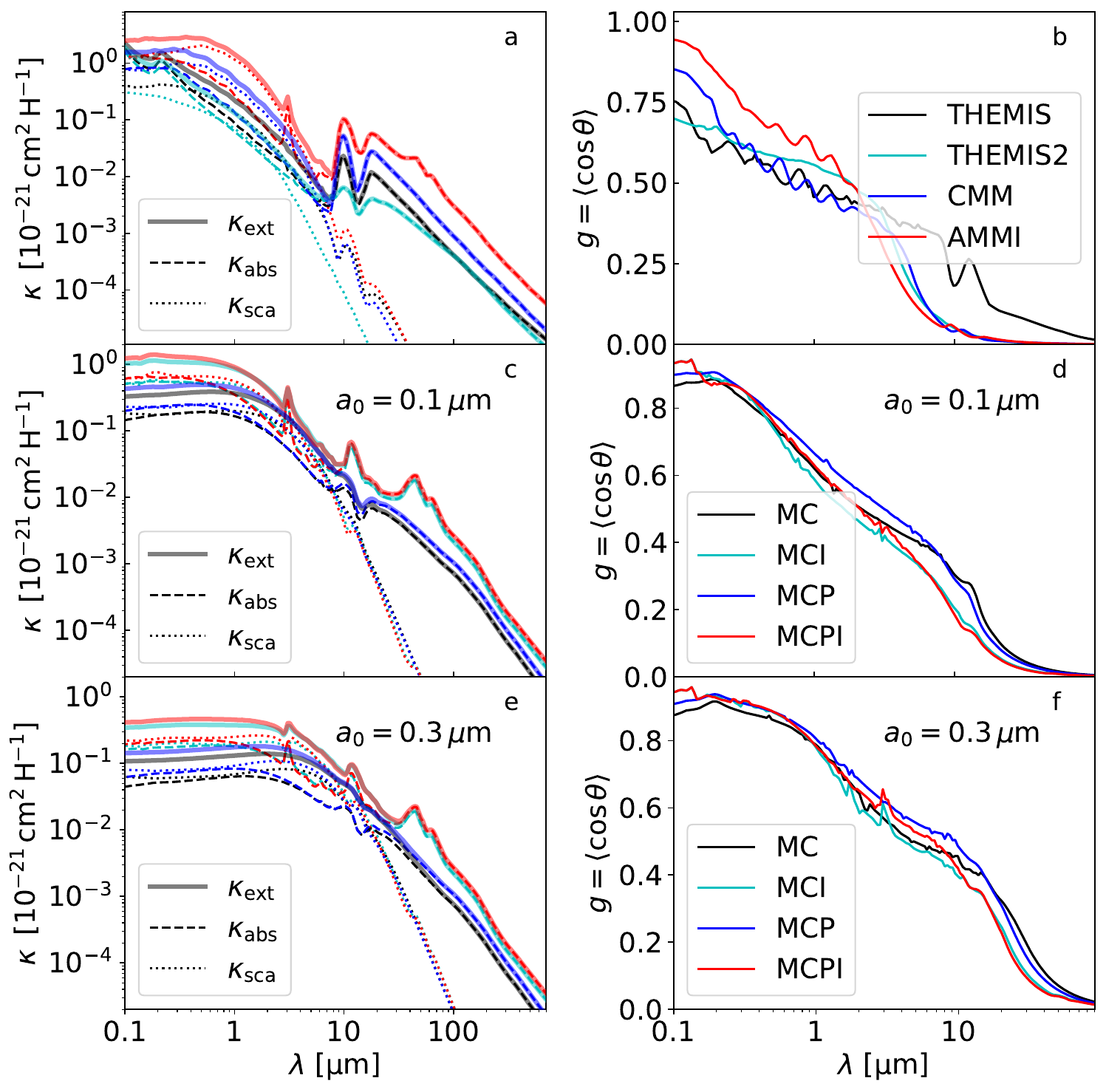}
  \caption{Dust cross-sections $\kappa_{\rm ext}$, $\kappa_{\rm abs}$, and $\kappa_{\rm sca}$
    per hydrogen atom (left frames) and asymmetry parameters $g=\langle \cos \theta \rangle$ (right
    frames) for selected dust models. The uppermost frames show the data for the THEMIS, THEMIS2,
    CMM, and AMMI models.  The other frames show the corresponding data for the MC, MCI, MCP, and
    MCPI models, for the $a_0=0.1\,\mu$m (frames c-d) and $a_0=0.3\,\mu$m (frames e-f) size
    distributions.  }
  \label{fig:Qg}
\end{figure}

The optical properties of most dust models can be found through the
DustEM\footnote{https://www.ias.u-psud.fr/DUSTEM/} and
THEMIS\footnote{https://www.ias.u-psud.fr/themis/THEMIS\_model.html} web sites and in the articles
listed in Sect~\ref{sect:dust}. The exceptions are MC, MCI, MCP, and MCPI that were calculated for
this paper. Figure~\ref{fig:Qg} shows examples of optical properties for these four models, 
  as well as for the THEMIS, THEMIS2, CMM, and AMMI models. In the plot, the scattering functions
$\phi(\theta)$ are characterised by the asymmetry parameters $g = \langle \cos \theta \rangle$,
where $\theta$ is the scattering angle. However, the model calculations used the effective
scattering function, average $\phi$ weighted by the scattering cross- sections of grains of
different sizes, rather than approximating the scattering function with a single Henyey-Greenstein
function and the similarly weighted average $g$ \citep[e.g.][]{Steinacker2013}.

\begin{figure*}
  \sidecaption
  \includegraphics[width=12cm]{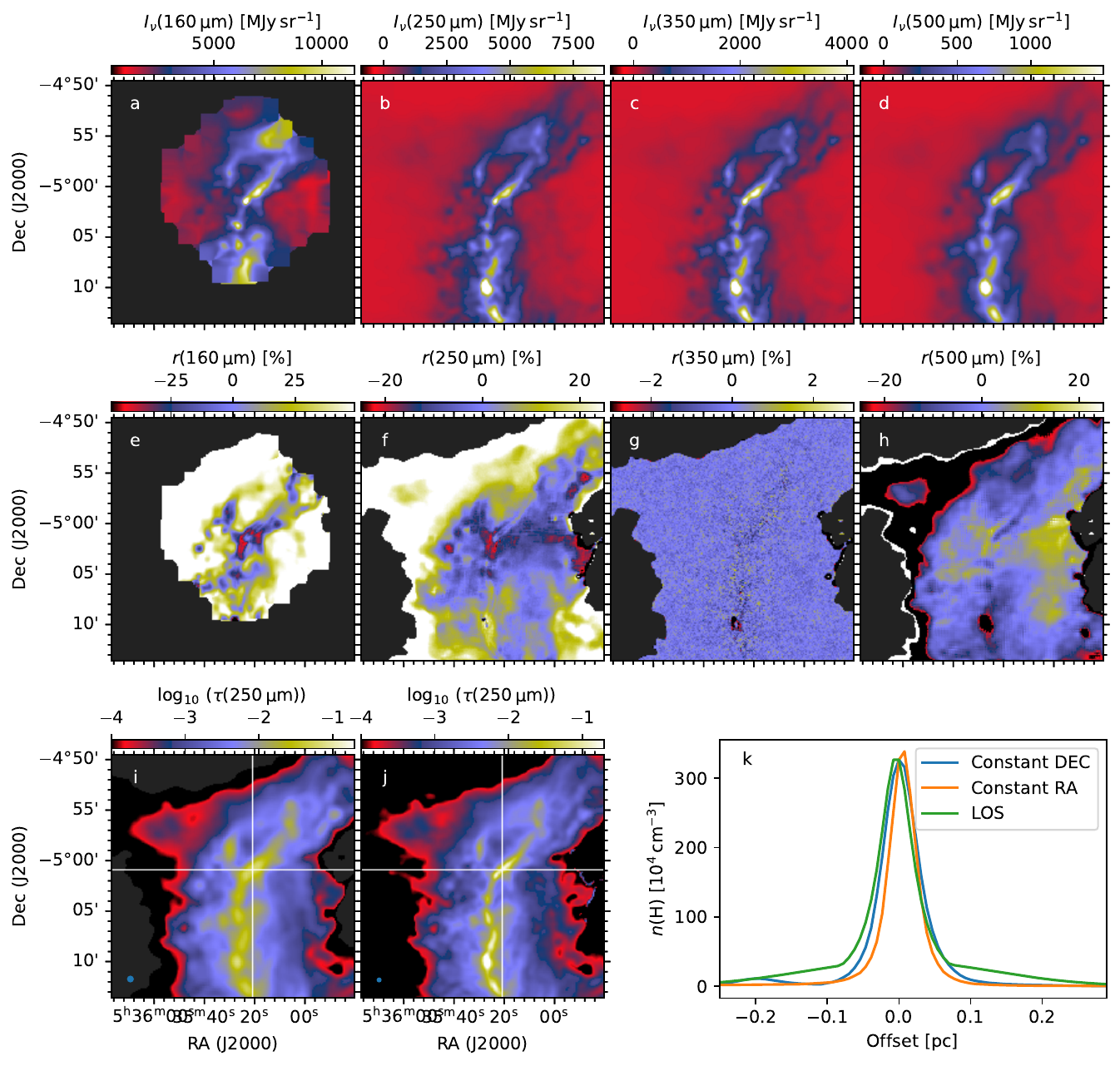}
  \caption{
    Results for cloud model B with $A_{\rm V}^{\rm ext}=1$\,mag and $a_0=0.3\,\mu$m Mix 1
    dust model. Frames a-d show the observed surface brightness maps and frames e-h the
    corresponding relative 160-500\,$\mu$m residuals, $(I_{\nu}^{\rm OBS}-I_{\nu}^{\rm MOD})/
    I_{\nu}^{\rm OBS}$.  Frame i shows a map of the 250\,$\mu$m optical depth from the direct MBB
    fitting of observations (41$\arcsec$ resolution), and frame j shows the same for the fitted
    model (30$\arcsec$ resolution).  The last frame shows one-dimensional cross-sections of the
    model densities along the lines marked with white lines in the previous frames, along the LOS
    central plane of the model.  }
  \label{fig:plot_examples_1}
\end{figure*}

As an example of fits with constant dust properties, Fig.~\ref{fig:plot_examples_1} shows the
residuals, optical depths, and density cross-sections for a run using Mix 1 dust with
$a_0=0.3\,\mu$m.  The results are shown for cloud model B and $A_{\rm V}^{\rm ext}=1$\,mag of
external extinction.

\begin{figure*}
  \includegraphics[width=18.6cm]{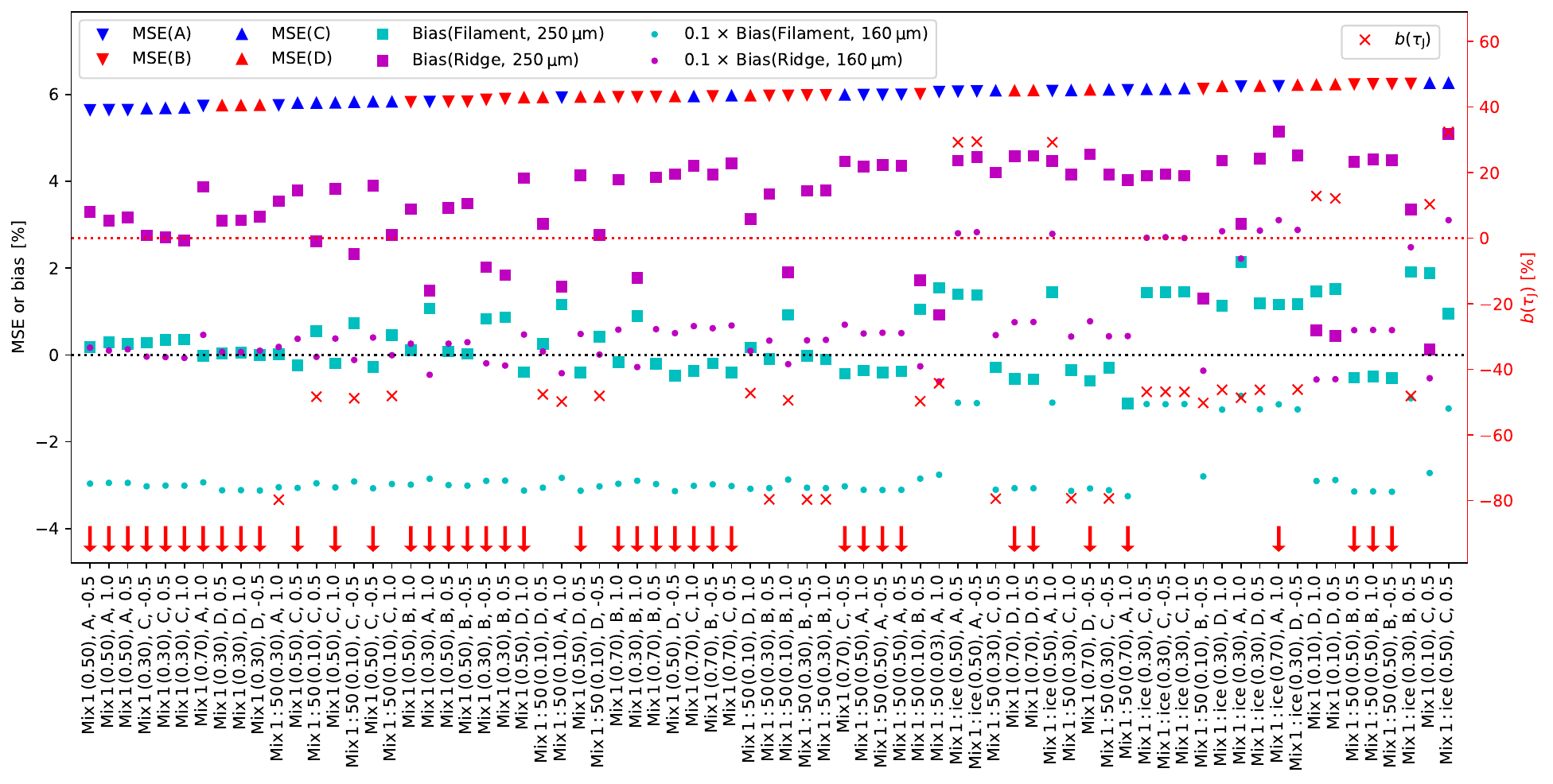}
  \caption{
    Best single-dust models in order of increasing MSE, based on average of 250\,$\mu$m
    and 500\,$\mu$m residuals in $Q=1-2$ areas. Each model is labelled by the combination of
    the dust and cloud names and the $A^{\rm ext}_{\rm V}$ value. The MSE markers and their colours
    indicate the model versions A-D. The 250\,$\mu$m and 160\,$\mu$m bias values are also
      plotted using the left y-axis, separately for the filament ($Q=1$) and the ridge ($Q=2$)
      regions. The 160\,$\mu$m bias values are scaled by 0.1 for the plotting.
    The red crosses and the right y-axis show the bias in the predicted J-band extinction,
    $b(\tau_{\rm J})= (A{\rm (J,model)}-A{\rm (J,obs.)})/A{\rm (J,obs.)}$ over the $Q=1$ area.
    $A{\rm (J, \,obs.)}$ is estimated using the extinction curve of the dust model in question and
    all stars in the filament region ($Q=1$). Red arrows indicate values outside the plotted range.
  }
  \label{fig:plot_single_3}
\end{figure*}

Figure~\ref{fig:plot_single_3} shows data for 80 models with spatially constant dust properties.
The cases are selected based on their lowest MSE values, which are obtained by taking the average
  of the 250\,$\mu$m and 500\,$\mu$m MSE values in the $Q=1$ and $Q=2$ areas.

\section{Dust models and NIR extinction} \label{sect:dust_table}

Table~\ref{table:dust_table} lists key parameters for the dust models tested. These include the
spectral index in the 250-500\,$\mu$m wavelength interval and the ratio of the J-band and
250\,$\mu$m dust opacities, $\tau(250\,\mu{\rm m})/\tau({\rm J})$. The table gives a summary of the
parameter combinations tested in the case of the single-dust models and indicates the cases showing
saturation (i.e. imperfect fit to the observed 350\,$\mu$m data). Only the case with $A_{\rm V}^{\rm
  ext}$=1\,mag was calculated for every combination of dust properties and A-D versions of
cloud LOS density profiles.
  
\begin{figure}
\includegraphics[width=8.8cm]{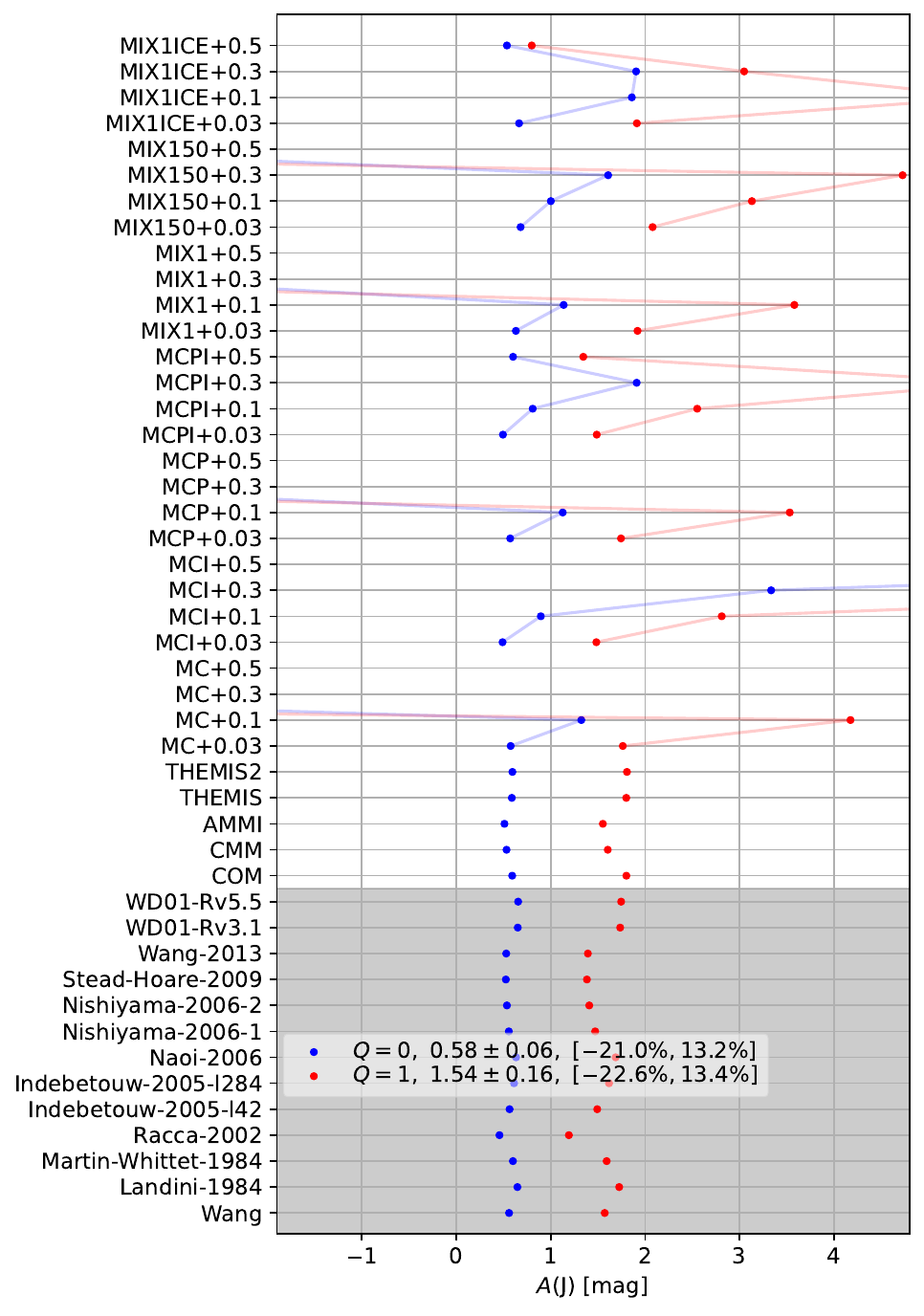}
\caption{
Mean extinction $A(J)$ in $Q=0$ (blue symbols) and $Q=1$ (red symbols) areas. Each row
corresponds to a different extinction curve. The bottom shaded area includes the different observed
extinction curves listed in Table 2 in \cite{Wang2014}, and the upper rows are a selection of the
dust models used in this paper. All extinction estimates are background-subtracted, and these can
diverge, even with negative values, for models with grains sizes $a_0 \gtrsim 0.1$\,$\mu$m.
}
\label{fig:plot_cases}
\end{figure}

In Fig.~\ref{fig:plot_cases} the shaded area shows extinction estimates for the set of extinction
curves listed in Table 5 of \cite{Wang2014}. These refer mainly to
observations at low column densities. However, one case has the ratio of total to selective
extinction $R_{\rm V}=5.5$, which should correspond to a dense environment with significant grain
growth \citep{WeingartnerDraine2001}. All these extinction curves result in less than 15\%
variation in the NIR extinction estimates. The \cite{Indebetouw2005} colour ratios used by
\citet{Meingast2016} are among this group.

Figure~\ref{fig:plot_cases} includes additional extinction estimates for a set of dust models used in
this paper. Models with mean grain sizes below $\sim$0.1\,$\mu$m give values similar to those
discussed above. However, larger grains result in rapid changes in the estimates. The values of the
$Q=1$ region first increase and, for grain sizes $a_0 \gtrsim 0.3$\,$\mu$m, can jump to even
negative values. This agrees qualitatively with behaviour seen in
Fig.~\ref{fig:plot_dusts_B}a. As the NIR extinction curve becomes flat, there is no longer a
clear correlation between the observed stellar colours and the extinction.

\begin{table*}
  \caption{FIR dust opacity spectral indices $\beta$ and FIR-to-NIR opacity ratios
    $\tau(250\,\mu{\rm m})/\tau({\rm J})$ for dust models used in modelling.  The last four
    columns list the $A_{\rm V}^{\rm ext}$ magnitude values tested for each of the cloud models
    A-D, using the dust model of the first column (single-dust models only). Red colour indicates
    cases with incomplete convergence, the 350\,$\mu$m MSE values exceeding 1\% in the $Q=2$
    region.  }
  \begin{center}
    \begin{tabular}{lcccccc}
      \hline
      Dust &   $\beta(250-500\,\mu {\rm m})$ &  $\tau(250\,\mu{\rm m})/\tau({\rm J})$  & A & B & C       & D \\
           &                                 &    $\rm [10^{-4}]$ & & & & \\
      \hline
                 COM  &    1.83  &   4.87  &  0.5, 1.0 &  -0.5, 0.5, 1.0 &  -0.5, 0.5, 1.0 &  -0.5, 0.5, 1.0 \\
                 CMM  &    2.03  &   3.93  &  0.5, 1.0 &  -0.5, \textcolor{red}{ 0.5}, 1.0 &  -0.5, 0.5, 1.0 &  -0.5, 0.5, 1.0 \\
                 AMM  &    1.98  &   4.07  &  0.5, 1.0 &  -0.5, 0.5, 1.0 &  -0.5, 0.5, 1.0 &  -0.5, 0.5, 1.0 \\
                AMMI  &    2.02  &   3.98  &  0.5, 1.0 &  -0.5, 0.5, 1.0 &  -0.5, 0.5, 1.0 &  -0.5, 0.5, 1.0 \\
              THEMIS  &    1.75  &   4.69  &  0.5, 1.0 &  -0.5, 0.5, 1.0 &  -0.5, 0.5, 1.0 &  -0.5, 0.5, 1.0 \\
             THEMIS2  &    1.92  &   7.08  &  0.5, 1.0 &  -0.5, 0.5, 1.0 &  -0.5, 0.5, 1.0 &  -0.5, 0.5, 1.0 \\
                  MC  &    2.22  &   2.61  & \textcolor{red}{ 0.5}, \textcolor{red}{ 1.0} &  0.5, \textcolor{red}{ 1.0} &  0.5, 1.0 & \textcolor{red}{ 0.5}, 1.0 \\
                 MCI  &    2.15  &   3.21  &  0.5, 1.0 & \textcolor{red}{ 0.5}, \textcolor{red}{ 1.0} &  0.5, 1.0 &  0.5, 1.0 \\
                 MCP  &    2.22  &   3.67  & \textcolor{red}{ 0.5}, 1.0 &  0.5, 1.0 &  0.5, 1.0 &  0.5, 1.0 \\
                MCPI  &    2.15  &   4.19  &  0.5, 1.0 &  0.5, \textcolor{red}{ 1.0} &  0.5, 1.0 &  0.5, 1.0 \\
         MC+0.03+1.0  &    2.22  &   3.16  & \textcolor{red}{ 1.0 } & \textcolor{red}{ -0.5}, 0.5, 1.0 & \textcolor{red}{ -0.5}, 0.5, 1.0 & \textcolor{red}{ -0.5}, 0.5, 1.0 \\
          MC+0.1+3.0  &    2.22  &   3.17  & \textcolor{red}{ 1.0 } & \textcolor{red}{ -0.5}, \textcolor{red}{ 0.5}, \textcolor{red}{ 1.0} & \textcolor{red}{ -0.5}, \textcolor{red}{ 0.5}, \textcolor{red}{ 1.0} & \textcolor{red}{ -0.5}, \textcolor{red}{ 0.5}, \textcolor{red}{ 1.0} \\
          MC+0.3+5.0  &    2.23  &   8.03  &  1.0  & \textcolor{red}{ -0.5}, \textcolor{red}{ 0.5}, \textcolor{red}{ 1.0} & \textcolor{red}{ -0.5}, \textcolor{red}{ 0.5}, \textcolor{red}{ 1.0} & \textcolor{red}{ -0.5}, \textcolor{red}{ 0.5}, \textcolor{red}{ 1.0} \\
          MC+0.5+5.0  &    2.24  &  12.99  &  -0.5, 0.5, 1.0 &  -0.5, 0.5, 1.0 &  -0.5, 0.5, 1.0 &  -0.5, 0.5, 1.0 \\
        MCI+0.03+1.0  &    2.15  &   4.45  &  1.0  &  -0.5, 0.5, 1.0 &  -0.5, 0.5, 1.0 &  -0.5, 0.5, 1.0 \\
         MCI+0.1+3.0  &    2.15  &   3.35  &  1.0  &  -0.5, 0.5, 1.0 & \textcolor{red}{ -0.5}, 0.5, 1.0 & \textcolor{red}{ -0.5}, 0.5, 1.0 \\
         MCI+0.3+5.0  &    2.15  &   6.67  &  1.0  & \textcolor{red}{ -0.5}, 0.5, 1.0 & \textcolor{red}{ -0.5}, \textcolor{red}{ 0.5}, \textcolor{red}{ 1.0} & \textcolor{red}{ -0.5}, \textcolor{red}{ 0.5}, \textcolor{red}{ 1.0} \\
         MCI+0.5+5.0  &    2.16  &  10.40  &  -0.5, 0.5, 1.0 &  -0.5, 0.5, 1.0 &  -0.5, 0.5, 1.0 &  -0.5, 0.5, 1.0 \\
        MCP+0.03+1.0  &    2.22  &   4.60  &  1.0  &  -0.5, 0.5, 1.0 & \textcolor{red}{ -0.5}, 0.5, 1.0 & \textcolor{red}{ -0.5}, 0.5, 1.0 \\
         MCP+0.1+3.0  &    2.22  &   3.99  & \textcolor{red}{ 1.0 } & \textcolor{red}{ -0.5}, 0.5, 1.0 & \textcolor{red}{ -0.5}, \textcolor{red}{ 0.5}, 1.0 & \textcolor{red}{ -0.5}, \textcolor{red}{ 0.5}, \textcolor{red}{ 1.0} \\
         MCP+0.3+5.0  &    2.23  &   8.93  &  1.0  & \textcolor{red}{ -0.5}, \textcolor{red}{ 0.5}, \textcolor{red}{ 1.0} & \textcolor{red}{ -0.5}, \textcolor{red}{ 0.5}, \textcolor{red}{ 1.0} & \textcolor{red}{ -0.5}, \textcolor{red}{ 0.5}, \textcolor{red}{ 1.0} \\
         MCP+0.5+5.0  &    2.23  &  14.19  &  -0.5, 0.5, 1.0 &  -0.5, 0.5, 1.0 &  -0.5, 0.5, 1.0 &  -0.5, 0.5, 1.0 \\
       MCPI+0.03+1.0  &    2.15  &   6.01  &  1.0  &  -0.5, 0.5, 1.0 &  -0.5, 0.5, 1.0 &  -0.5, 0.5, 1.0 \\
        MCPI+0.1+3.0  &    2.15  &   4.02  &  1.0  &  -0.5, 0.5, 1.0 &  -0.5, 0.5, 1.0 & \textcolor{red}{ -0.5}, 0.5, 1.0 \\
        MCPI+0.3+5.0  &    2.15  &   6.91  &  1.0  &  -0.5, 0.5, 1.0 & \textcolor{red}{ -0.5}, 0.5, 1.0 & \textcolor{red}{ -0.5}, 0.5, 1.0 \\
        MCPI+0.5+5.0  &    2.16  &  10.55  &  -0.5, 0.5, 1.0 &  -0.5, 0.5, 1.0 &  -0.5, 0.5, 1.0 &  -0.5, 0.5, 1.0 \\
       MIX 1+0.03+1.0  &    1.52  &   5.26  &  1.0  &  -   &  -   &  -   \\
        MIX 1+0.1+3.0  &    1.53  &   4.69  &  1.0  &  -0.5, 0.5, 1.0 &  -0.5, 0.5, 1.0 &  -0.5, 0.5, 1.0 \\
       MIX 1+0.3+10.0  &    1.55  &  11.52  &  1.0  &  -0.5, 0.5, 1.0 &  -0.5, 0.5, 1.0 &  -0.5, 0.5, 1.0 \\
       MIX 1+0.5+10.0  &    1.58  &  20.83  &  -0.5, 0.5, 1.0 &  -0.5, 0.5, 1.0 &  -0.5, 0.5, 1.0 &  -0.5, 0.5, 1.0 \\
     MIX150+0.03+1.0  &    1.50  &  15.34  &  1.0  &  -   &  -   &  -   \\
      MIX150+0.1+3.0  &    1.50  &  11.54  &  1.0  &  -0.5, 0.5, 1.0 &  -0.5, 0.5, 1.0 &  -0.5, 0.5, 1.0 \\
     MIX150+0.3+10.0  &    1.51  &  20.26  &  1.0  &  -0.5, 0.5, 1.0 &  -0.5, 0.5, 1.0 &  -0.5, 0.5, 1.0 \\
     MIX150+0.5+10.0  &    1.52  &  33.35  &  -0.5, 0.5, 1.0 &  -0.5, 0.5, 1.0 &  -0.5, 0.5, 1.0 &  -0.5, 0.5, 1.0 \\
    MIX1ICE+0.03+1.0  &    1.58  &   6.90  &  1.0  &  -   &  -   &  -   \\
     MIX1ICE+0.1+3.0  &    1.58  &   6.36  &  1.0  &  -0.5, 0.5, 1.0 &  -0.5, 0.5, 1.0 &  -0.5, 0.5, 1.0 \\
    MIX1ICE+0.3+10.0  &    1.60  &  14.80  &  1.0  &  -0.5, 0.5, 1.0 &  -0.5, 0.5, 1.0 &  -0.5, 0.5, 1.0 \\
    MIX1ICE+0.5+10.0  &    1.65  &  24.85  &  -0.5, 0.5, 1.0 &  -0.5, 0.5, 1.0 &  -0.5, 0.5, 1.0 &  -0.5, 0.5, 1.0 \end{tabular}
\end{center}
\label{table:dust_table}
\end{table*}

\end{appendix}

\end{document}